\documentclass[aps,prd,twocolumn,superscriptaddress,showpacs,preprintnumbers,nofootinbib,altaffilletter,bibnotes]{revtex4}

\usepackage{graphicx}

\begin{document}

\preprint{Submitted to Phys. Rev. D}

\title{Search for Neutrino-Induced Cascades with the AMANDA Detector}

\author{J.~Ahrens}
\affiliation{Institute of Physics, University of Mainz, Staudinger Weg 7, D-55099 Mainz, Germany}
\author{X.~Bai}
\affiliation{Bartol Research Institute, University of Delaware, Newark, DE 19716, USA}
\author{G.~Barouch}
\affiliation{Dept. of Physics, University of Wisconsin, Madison, WI 53706, USA}
\author{S.W.~Barwick}
\affiliation{Dept. of Physics and Astronomy, University of California, Irvine, CA 92697, USA}
\author{R.C.~Bay}
\affiliation{Dept. of Physics, University of California, Berkeley, CA 94720, USA}
\author{T.~Becka}
\affiliation{Institute of Physics, University of Mainz, Staudinger Weg 7, D-55099 Mainz, Germany} 
\author{K.-H.~Becker}
\affiliation{Fachbereich 8 Physik, BUGH Wuppertal, D-42097 Wuppertal, Germany}
\author{D.~Bertrand}
\affiliation{Universit\'e Libre de Bruxelles, Science Faculty CP230, Boulevard du Triomphe, B-1050 Brussels, Belgium}
\author{F.~Binon}
\affiliation{Universit\'e Libre de Bruxelles, Science Faculty CP230, Boulevard du Triomphe, B-1050 Brussels, Belgium}
\author{A.~Biron}
\affiliation{DESY-Zeuthen, D-15735 Zeuthen, Germany}
% ...
\collaboration{The AMANDA Collaboration}
\noaffiliation

\date{\today}

\begin{abstract}

\noindent We report on a search for electromagnetic and/or hadronic showers
(\textit{cascades}) induced by a diffuse flux of 
neutrinos with energies between 5~TeV and 300~TeV
from extraterrestrial sources.  Cascades may be produced by matter
interactions of all flavors of neutrinos, and contained cascades have
better energy resolution and afford better background rejection than
through-going $\nu_\mu$-induced muons.  Data taken in 1997 with the
AMANDA detector were searched for events with a high-energy cascade-like
signature. The observed events are consistent with expected
backgrounds from atmospheric neutrinos and catastrophic energy losses
from atmospheric muons. Effective volumes for all flavors of
neutrinos, which allow the calculation of limits for any neutrino flux
model, are presented. The limit on cascades from a diffuse flux of
$\nu_e+\nu_\mu+\nu_\tau+\overline{\nu}_e+\overline{\nu}_\mu+\overline{\nu}_\tau$
is $E^2 \frac{d\Phi}{dE} < 9.8 \times
10^{-6}\;\mathrm{GeV\,cm^{-2}\,s^{-1}\,sr^{-1}}$, assuming a neutrino
flavor flux ratio of 1:1:1 at the detector.  The limit on cascades
from a diffuse flux of $\nu_e+\overline{\nu}_e$ is $E^2 \frac{d\Phi}{dE} < 6.5
\times 10^{-6}\;\mathrm{ GeV\,cm^{-2}\,s^{-1}\,sr^{-1}}$, independent 
of the assumed neutrino flavor flux ratio.

\end{abstract} 


\maketitle

\section{Introduction}

Neutrinos interact principally via the weak force, posing a detection
challenge for neutrino telescopes but bestowing a valuable advantage
on the field of neutrino astronomy: neutrino fluxes from astronomical
sources are essentially unattenuated even over cosmological
distances. In contrast, high-energy gamma rays are absorbed and/or
scattered by intervening matter and photons, and high-energy
cosmic-rays are deflected by galactic and intergalactic magnetic
fields except at the highest energies ($>10^{19}$~eV).

We present a search for the fully-reconstructed light patterns created
by electromagnetic or hadronic showers (\textit{cascades}) resulting
from a diffuse flux of high-energy 
extraterrestrial neutrinos.  We use data collected in 1997 from the
Antarctic Muon and Neutrino Detector Array (AMANDA) for this purpose.
Demonstrating $\nu$-induced cascade sensitivity is an important step
for neutrino astronomy because the cascade channel probes all neutrino
flavors, whereas the muon channel is primarily sensitive to charged
current $\nu_\mu$ and $\overline{\nu}_\mu$ interactions.  This is
particularly relevant in view of the emerging understanding of
neutrino oscillations~\cite{sno-cc,sno-nc,sno-dn,superk-atm}, in which
the flux of $\nu_\mu$ would be reduced by oscillations.  (The
detection of high-energy atmospheric muon neutrinos by AMANDA has been
demonstrated by the full reconstruction of Cherenkov light patterns
produced by up-going
muons~\cite{ama:nature-pub,amanda:atm-b10,amanda:wimps-b10}.)
Cascades also boast more accurate energy measurement and better
separation from background, although they suffer from worse angular
resolution and reduced effective volume relative to muons.
Importantly, it is straightforward to calibrate the cascade response
of neutrino telescopes such as AMANDA at lower energies through use
of, e.g., \textit{in-situ} light sources.  Furthermore, cascades become
increasingly easier to identify and reconstruct as detector volumes
get larger, so the techniques presented here have relevance for future
analyses performed at larger detectors.

Electron neutrinos can produce cascades with no
detectable track via the charged current (CC) interaction and all
neutrino flavors can produce cascades via the neutral current
(NC) interaction.  Cascade-like events are also produced in $\nu_\tau$
CC interactions when the resulting $\tau$ decays into an electron
(roughly 18\% branching ratio) or into mesons (roughly 64\% branching
ratio) and the $\tau$ energy is below about 100~TeV, at which energy
the $\tau$ decay length is less than 5~m, so that the shower produced
by the neutrino interaction and the shower produced by the $\tau$
decay cannot be spatially resolved by AMANDA. The contribution of
$\nu_\tau$ to the cascade channel becomes important when flavor
oscillations are taken into account for
extraterrestrial~\cite{halzen-saltzberg,reno,beacom} and for
atmospheric~\cite{stanev:prl} $\nu$-induced cascades. For
extraterrestrial sources, current knowledge of neutrino oscillations
suggests a detected neutrino flavor flux ratio of
$\nu_e$:$\nu_\mu$:$\nu_\tau$::1:1:1 following an expected flux ratio
of 1:2:0 at the source.

The total light output of an electromagnetic cascade is approximately
$10^{8}$ photons/TeV in ice. Hadronic cascades have a light yield
about 20\% lower~\cite{wiebusch:phd}. An electromagnetic cascade
develops in a cylinder of about 10-15~cm in radius (\textit{Moli\`ere
radius}) and several meters in length (about 8.5~m from the vertex of
a 100~TeV cascade, essentially all charged particles are below the
critical energy).  Hadronic cascades have longer longitudinal
developments and larger Moli\`ere radii. As a sparsely instrumented
detector, AMANDA is insensitive to the topological differences between
electromagnetic and hadronic cascades.  Since the NC interaction has a
lower cross section and results in a deposition of less energy than
the CC interaction, and since we assume a steeply falling neutrino
energy spectrum, at any given energy a very small fraction of the
$\nu_e$ events are due to NC interactions.  Hence their impact on the
cascade energy resolution is small, and the energy spectrum of
reconstructed cascades closely follows that of the CC $\nu_e$ energy
spectrum. 

In this paper, we present limits on the diffuse fluxes of
($\nu_e+\nu_\mu+\nu_\tau+\overline{\nu}_e+\overline{\nu}_\mu+\overline{\nu}_\tau$)
and ($\nu_e+\overline{\nu}_e$), assuming a customary $E^{-2}$ power law spectrum at
the source.  These limits are based on the observation of no events
consistent with a diffuse flux of high-energy extraterrestrial
neutrinos. We also present effective volumes for all neutrino flavors
to facilitate the calculation of a limit for any flux model.  (A search
for up-going muons produced by a $\nu_\mu$ extraterrestrial diffuse
flux is presently being conducted and preliminary results have been
reported in~\cite{icrc01:dif}.)

\section{The AMANDA-B10 Detector}

The data used in this work was taken with the AMANDA-B10 detector in
1997. AMANDA-B10~\cite{amanda:atm-b10,amanda:wimps-b10,icrc99:b10}
was commissioned in 1997 with a total of 302 optical modules (OMs)
arranged on 10 strings, at depths between 1500~m and 2000~m below the
surface of the ice at the South Pole. The strings are arranged in two
concentric circles 35~m and 60~m in radius, with one string at the
center. The OMs in the inner four (outer six) strings have a 20~m
(10~m) vertical separation.  Each OM contains a 20~cm photo-multiplier
tube (PMT) in a spherical pressure vessel. Coaxial cables in the inner
four strings and twisted pair cables in the outer six strings provide
high voltage to the PMTs and simultaneously transmit their signals to
the electronics housed on the surface. The detector is triggered using
a majority condition in which an event is recorded if more than 16
modules have a signal (i.e., were ``hit'') in a 2~$\mu$s time window.
A total of $1.05 \times 10^9$ events were recorded during an effective
live-time of 130.1 days.

The optical properties of the ice have been studied with \textit{in-situ}
light sources and with atmospheric muons. These studies have shown
that ice at the South Pole is not perfectly homogeneous, but rather
consists of horizontal layers corresponding to global climatological
conditions in the past, such as ice ages. These layers lead to a
modulation of the absorption and effective scattering lengths as a
function of depth~\cite{price:geophys}.  Optical properties are also
modified by the presence of drill-hole bubbles which are created
during the drilling and deployment processes.

\section{\label{reco}Methods for Cascade Reconstruction}

Simple reconstruction algorithms are initially applied to the data.
These methods are used to reduce the data sample size and to seed more
sophisticated reconstruction algorithms, while maintaining high
passing rates for simulated signal events. For cascades, the mean
position of the hit OMs, or \textit{center of gravity}, is used as the
first guess of the position. In order to efficiently reject muons,
they too are reconstructed, beginning with a first guess track fit
called the \textit{line fit}~\cite{stenger}.  The line fit is an
algorithm that assumes that hits can be projected onto a line, and
that the particle which produced the hits travels with a velocity
$\vec{v}_{\rm line}$ and has a starting point $\vec{r}_0$. The fit
minimizes the quantity $\sum_{\rm i=1}^{{\rm N}_{\rm hits}}
(\vec{r}_{i}-\vec{r}_0-\vec{v}_{\rm line} \cdot t_i)^2$ as a function
of $\vec{r}_0$ and $\vec{v}_{\rm line}$, where ${\rm N}_{\rm hits}$ is
the number of hits in the event. These procedures are described in
more detail elsewhere~\cite{muon-reco,picrc}.

After calculating the first guesses, three maximum likelihood methods
are used consecutively to reconstruct precisely the cascade vertex
position, time, energy and direction. These methods are described
below.

\subsection{\label{pandel}Single Photoelectron Vertex Position and Time
Reconstruction}

The cascade vertex position and creation time are reconstructed using
a maximum likelihood function that takes into account the Cherenkov
emission, absorption and scattering of light.  This vertex information
is required for rejecting potential backgrounds and for subsequent
fits for energy and direction.  This procedure is quite similar to the
algorithms used for muon fitting~\cite{muon-reco}. A more
comprehensive description of the different cascade reconstruction
methods can be found in~\cite{picrc,kowalksi:diplm,taboada:phd}.

We use a likelihood function having the form:
\begin{equation}
\label{eq:pandel}
\mathcal{L}_{\vec{x},t}^{\rm spe} = \prod_{i=0}^{N_{\rm hits}} p(t_{\rm
res}^{i},d_i), 
\end{equation}
\noindent where $t_{\rm res} = t_{\rm hit} - t_{\rm Cher}$ is
the difference between observed hit time and expected time for
Cherenkov emission without scattering--the \textit{time residual}--and
$p(t_{\rm res},d)$ is the probability of observing a photon at a time
residual $t_{\rm res}$ at a distance $d$ from the emitter. The label
``spe'' indicates that $\mathcal{L}_{\vec{x},t}^{\rm spe}$
assumes all hits are due to single photoelectrons.

The probability $p(t_{\rm res},d)$ was generated by parametrizing
simulations of light propagation in ice. The product in
Eq.~\ref{eq:pandel} is calculated using all hit OMs. The maximization
of $\mathcal{L}_{\vec{x},t}^{\rm spe}$ provides a good estimate of the
vertex position and time of the cascade.

\subsection{\label{mpe} Multi-photoelectron Vertex Position and Time Reconstruction}

The single photoelectron likelihood can be refined by taking into
account that the time measured in each PMT is the time of the first
photon to be observed. If a PMT receives $N$ photons, the probability
of measuring a time residual, $t_{\rm res}$, and the associated
likelihood function are:

\begin{equation}
p(N,t_{\rm res},d) = N p(t_{\rm res},d) \left( \int_{t_{\rm res}}^{\infty} dt'
p(t',d) \right)^{N-1} 
\end{equation}
\begin{equation}
\label{eq:mpe}
\mathcal{L}_{\vec{x},t}^{\rm mpe} =
\prod_{i=0}^{N_{\rm hits}} p(N,t_{\rm res}^{i},d_i)
\end{equation}
where the ``mpe'' label indicates that $\mathcal{L}_{\vec{x},t}^{\rm
mpe}$ describes multi-photoelectron hits.  In AMANDA we use the
measured pulse amplitude as an estimator for $N$.

The maximization of $\mathcal{L}_{\vec{x},t}^{\rm mpe}$ is used to estimate
the most likely vertex position $\vec{x}$ and time $t$ of a cascade.  The
multi-photoelectron vertex reconstruction uses the maximization of
$\mathcal{L}_{\vec{x},t}^{\rm spe}$ to seed initial values of cascade
vertex position and time.

\subsection{\label{phit}Energy and Direction Reconstruction}

Cascade energy and direction are reconstructed using a 
likelihood function assembled from the probabilities of an OM
being hit or remaining unhit assuming a cascade hypothesis:

\begin{eqnarray}
\label{eq:phit} 
\mathcal{L}_{E,\hat{n}} = \prod^{\rm Hit\,OMs}_{i=0} P_{\rm
Hit}(d_i,E,\hat{n})\; \times \nonumber \\
\prod^{\rm Unhit\,OMs}_{i=0} P_{\rm
Unhit}(d_i,E,\hat{n})
\end{eqnarray}

\noindent Maximization of $\mathcal{L}_{E,\hat{n}}$ provides the most likely
value of energy $E$ and direction $\hat{n}$ of a cascade. Note that in principle
this procedure also allows for the reconstruction of position (but not
time) of a cascade. Monte Carlo studies have shown, however, that the
position resolution obtained by maximizing $\mathcal{L}_{E,\hat{n}}$
is not as good as that obtained with $\mathcal{L}_{\vec{x},t}^{\rm
spe}$ or $\mathcal{L}_{\vec{x},t}^{\rm mpe}$.

\section{Detector Response to Cascades}

\begin{figure}
\mbox{
\includegraphics[width=0.23\textwidth]{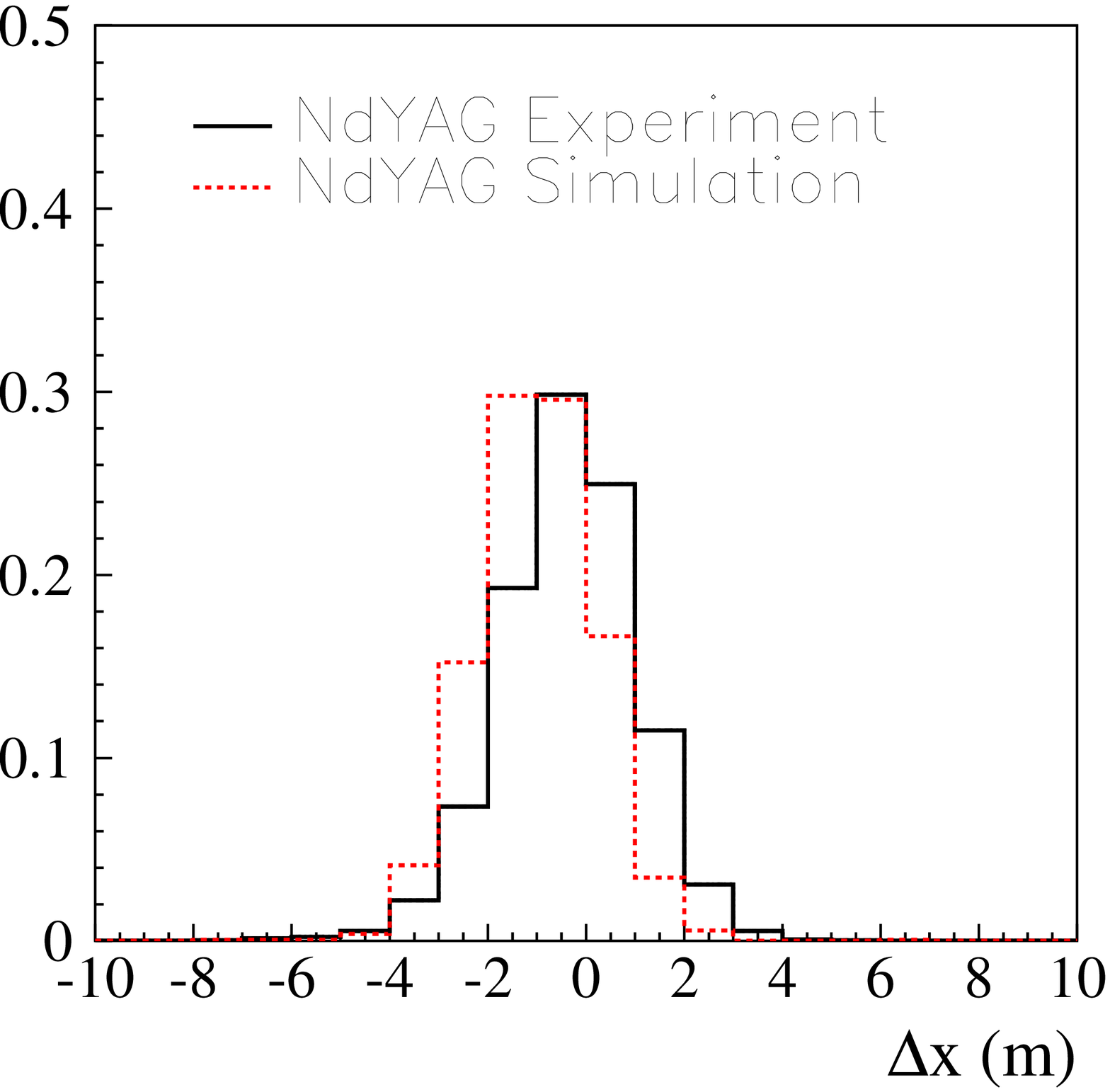}
\includegraphics[width=0.23\textwidth]{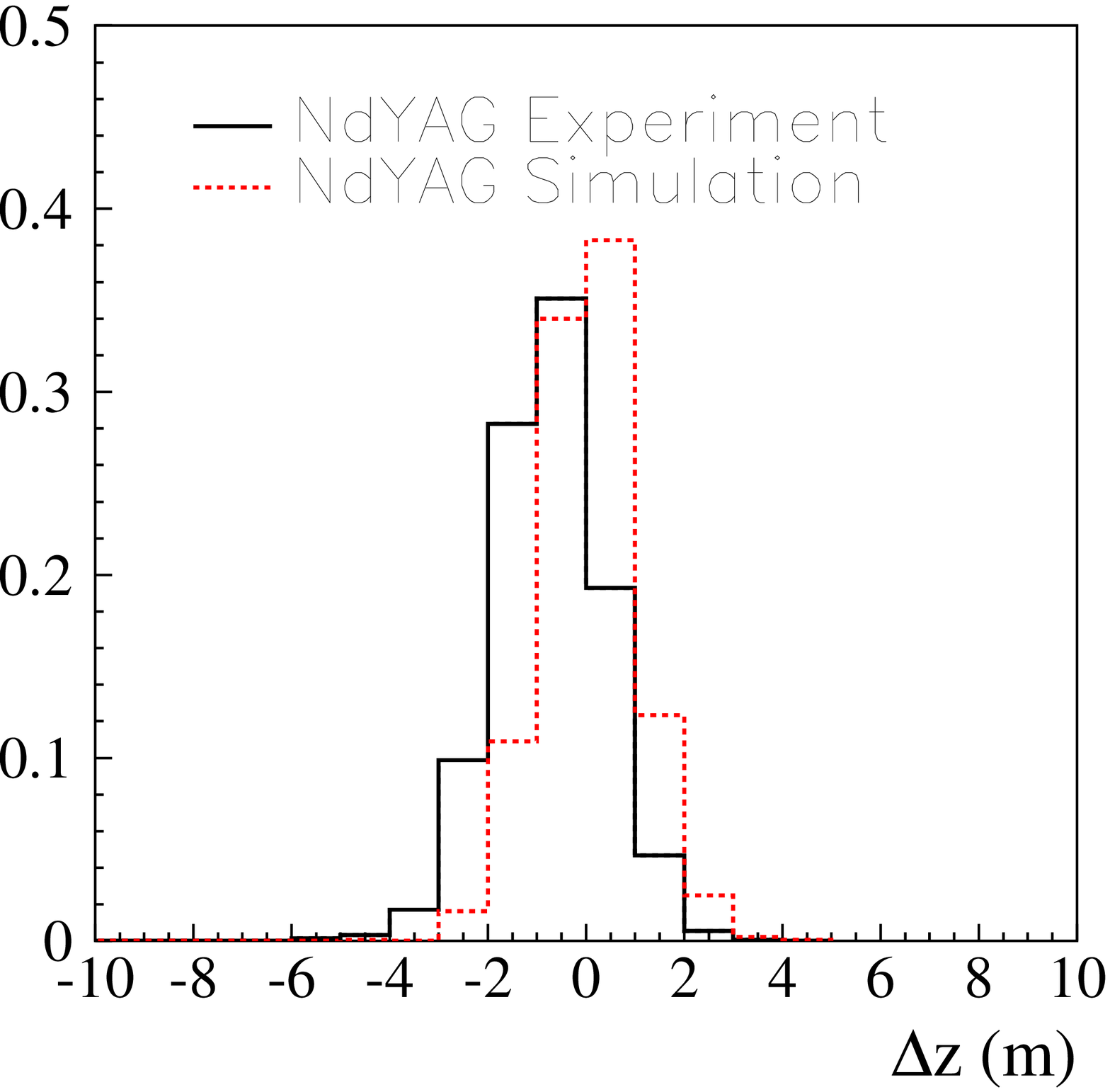}}
\mbox{
\includegraphics[width=0.23\textwidth]{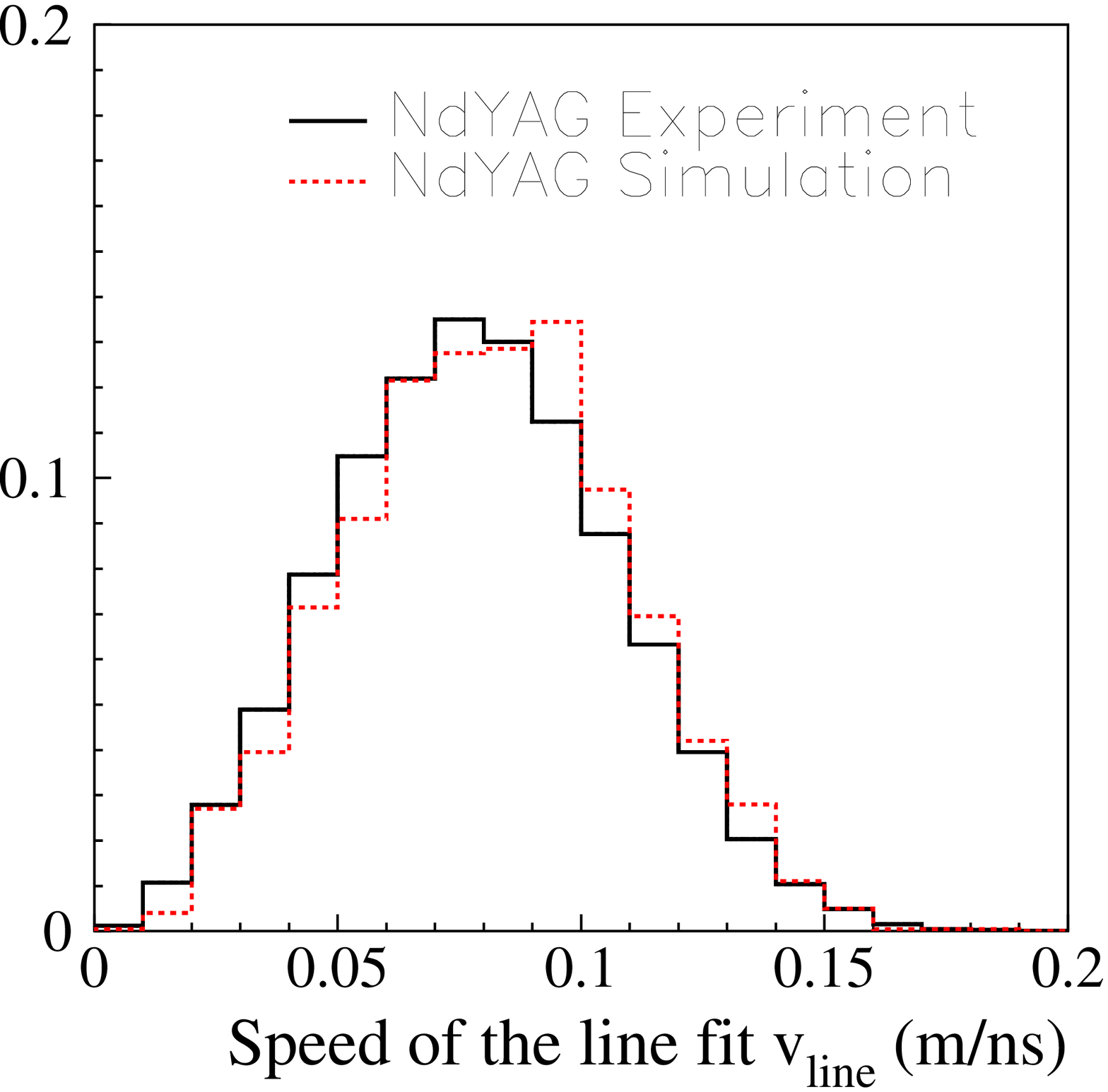}
\includegraphics[width=0.23\textwidth]{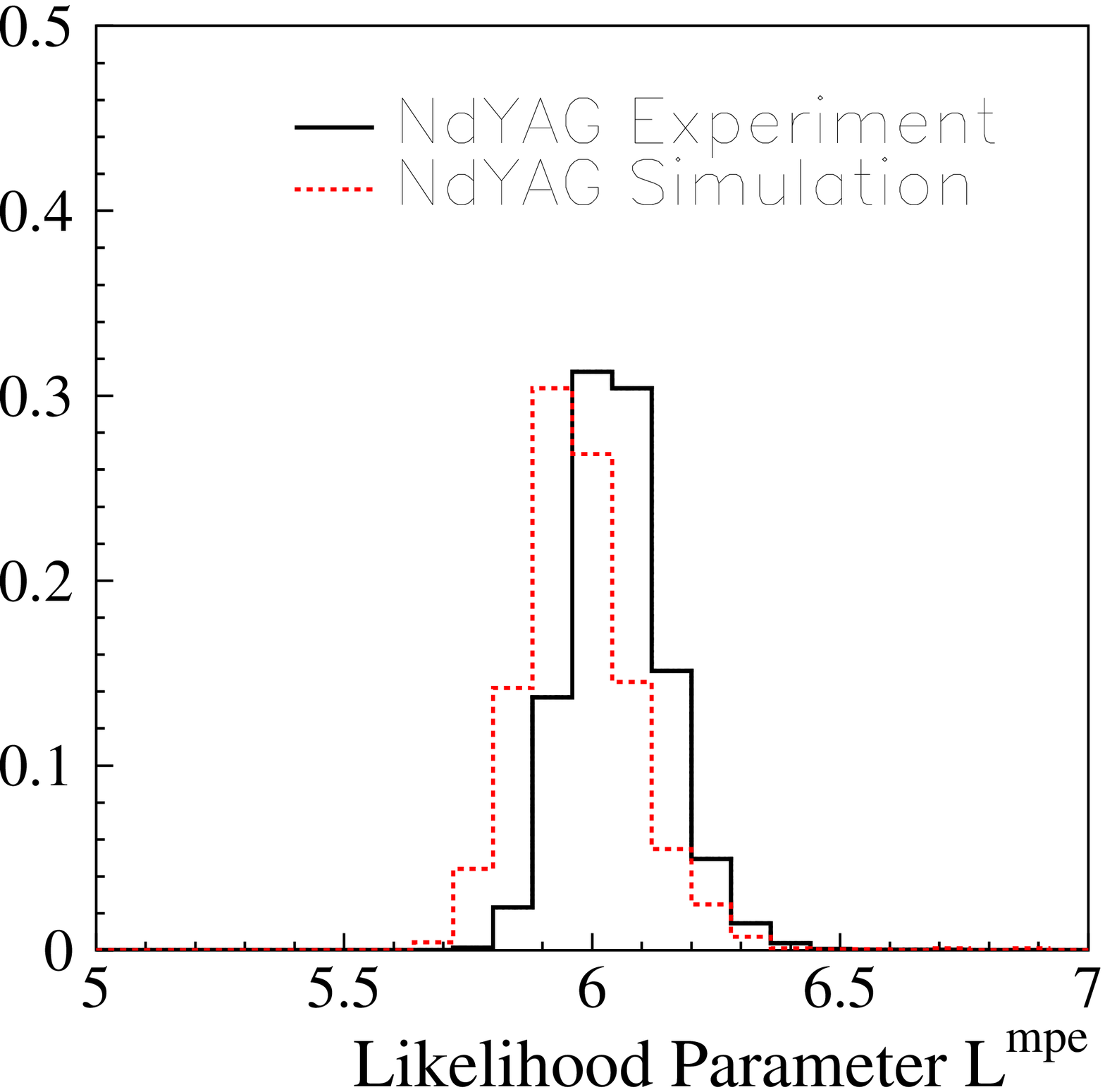}}
\caption{\label{z-reco-yag}
The plots show results of the reconstruction of pulsed laser data
(solid) and simulation (dashed) with the diffuser ball next to OM~69
(situated near the center of the detector). The differences between
simulated and reconstructed $x$ and $z$ components of the position are
shown on top. The speed of the line fit, $v_{\rm line}$, is shown in
the bottom left. The bottom right plot shows the reduced
$L_{\vec{x},t}^{\rm mpe}$, described in the text. The vertical scale
for all four plots is arbitrary. The position of the OMs
is known with about 1~m precision \cite{ama:b4-pub}, so there is no
discrepancy in the plot in the upper right. The discrepancy in the
plot in the lower right arises due to the simplified ice model used
for the pulsed Laser.}

\end{figure}

In the absence of a tagged source of high-energy neutrino-induced
cascades, to understand the response of the detector we rely on 
\textit{in-situ} light sources, catastrophic energy losses by down-going
cosmic-ray muons, and Monte Carlo simulations.  The successful
reconstruction of these data demonstrate detector sensitivity to
cascade signals.

\subsection{Pulsed Laser}

A pulsed laser operating at 532~nm on the surface is used to send
light through optical fibers to diffuser balls embedded in the ice
close to almost every OM in the detector. A comparison of Monte Carlo
and experimental data for these \textit{in-situ} light sources deepens
our understanding of reconstruction performance and detector signal
sensitivity. The photon intensity that can be produced at each
diffuser ball is not known \textit{a priori}, so we force the number
of hit channels in experimental and simulated pulsed laser data to
match. Thus, the simulations predict that the laser produces pulses in
the ice comprising $5 \times 10^7 - 1 \times 10^9$ photons
(corresponding to a maximum cascade energy of roughly 10~TeV). The
laser pulses are roughly 10~ns wide, short enough to mimic the time
structure of true cascades.  

Although highly useful as a cascade calibration source, the pulsed
laser system has some minor drawbacks.  The diffuser ball light output
is expected to be isotropic, so the laser data does not provide
information about the angular response of the detector to cascades.
The laser produces light at $\lambda=532$~nm and at this wavelength
the optical ice properties are different from those at Cherenkov
radiation wavelengths.  The effective scattering length at 532~nm is
18-30~m and depends on depth. The absorption length at 532~nm is 25~m
and independent of depth~\cite{price:geophys} (at the shorter
wavelengths characteristic of Cherenkov radiation the absorption
length is about 100~m).

Independent data sets taken with diffuser balls in a variety of
locations are reconstructed with the first guesses and with the
time-position reconstruction algorithm described above. The position
resolution is about 1~m in the $z$ dimension and about 2~m the
in $x$ and $y$.  It is better in $z$ due to closer OM spacing in
that dimension. 

The pulsed laser simulation uses a simplified optical model of ice
properties: the drill-hole bubbles are taken into account, but no
depth dependence is used for the scattering length. In spite of this
simplification, the vertex resolution of the pulsed laser data agrees
well with simulations, showing that the detector can be used to
reconstruct the position of contained point-like events.  (Contained
events are defined as events whose reconstructed vertex lies within a
right cylinder of height 400~m and radius 60~m, centered on the
AMANDA-B10 detector and encapsulated by it.)  Figure~\ref{z-reco-yag}
shows the results of the reconstruction of pulsed laser data and
simulation.

\subsection{\label{brem}Catastrophic Muon Energy Losses}

The vast majority of the events recorded by AMANDA are down-going
muons induced by cosmic-ray air showers. This background has been
simulated with the \texttt{Corsika} program~\cite{corsika} using the
average winter air density profile at the South Pole and the
\texttt{QGSJET} hadronic model~\cite{Kalmykov} option. The cosmic-ray
composition is taken from~\cite{cr-composition}. The propagation of
muons through ice is simulated with the program
\texttt{Mudedx}~\cite{lohmann-85,lohmann-95}. Optical properties of the ice,
including depth dependence and drill-hole bubbles, are also simulated.

From the large sample of atmospheric muons it is possible to extract a
subset of cascade-like events in which the majority of the recorded
hits come from catastrophic, localized energy loss of the muon (e.g.,
a bright bremsstrahlung).  The extraction of these events is achieved
using criteria 1-9 from table~\ref{cut_table}, i.e., we do not require
these events to reconstruct as up-going cascades.  (We do, however,
still reject obvious down-going muons via criterion number 8.)  Based
on the number of hits produced by the brightest cascade in simulated
events, and on a visual study of these events, we have confirmed that
after applying these selection criteria the remaining events are
indeed cascade-like.

\begin{figure}
\includegraphics[width=0.46\textwidth]{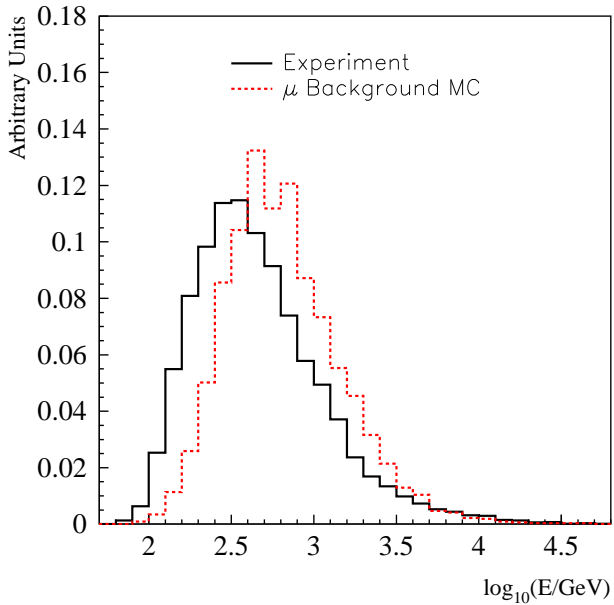}
\caption{ \label{fig:brem} Energy spectra of reconstructed
atmospheric muon energy losses for experimental data and standard simulated
muon background after application of selection criteria 1-9 from
table~\ref{cut_table}.  Agreement between simulation and experimental
data depends on a combination of the simulated ice properties and the
OM angular and absolute sensitivity. These effects have
been taken into account in the calculation of systematic uncertainties in
Section~\ref{systematics}.}
\end{figure}

Figure~\ref{fig:brem} shows the energy spectra of muon energy losses
for experimental data and simulated atmospheric muons. The
experimental and simulated data agree reasonably well, but not
perfectly. The difference is discussed in Section~\ref{systematics}.

\subsection{\label{mc_pred}Monte Carlo Prediction for Neutrino-Induced
Cascade Reconstruction}

To study the performance of the reconstruction algorithms we simulated
a flux of $\nu_e + \overline{\nu}_e$ following an $E^{-2}$ power law
spectrum. Neutrinos from astrophysical sources are expected to have a
hard spectrum, reflecting the processes in the cosmic accelerators
that generate them.  Earth absorption and NC scattering are taken into
account in the simulation. The ``Preliminary Reference Earth Model''
is used to calculate the Earth's density profile~\cite{earth}. We
calculate differential cross sections using \texttt{CTEQ5} following
Gandhi \textit{et al.}~\cite{gandhi}. For $\nu_\tau$ interactions, the
simulation of the $\tau$ decay uses
\texttt{TAUOLA}~\cite{tauola1,tauola2,tauola3}.

\begin{figure}[t]
\mbox{
\includegraphics[width=0.23\textwidth]{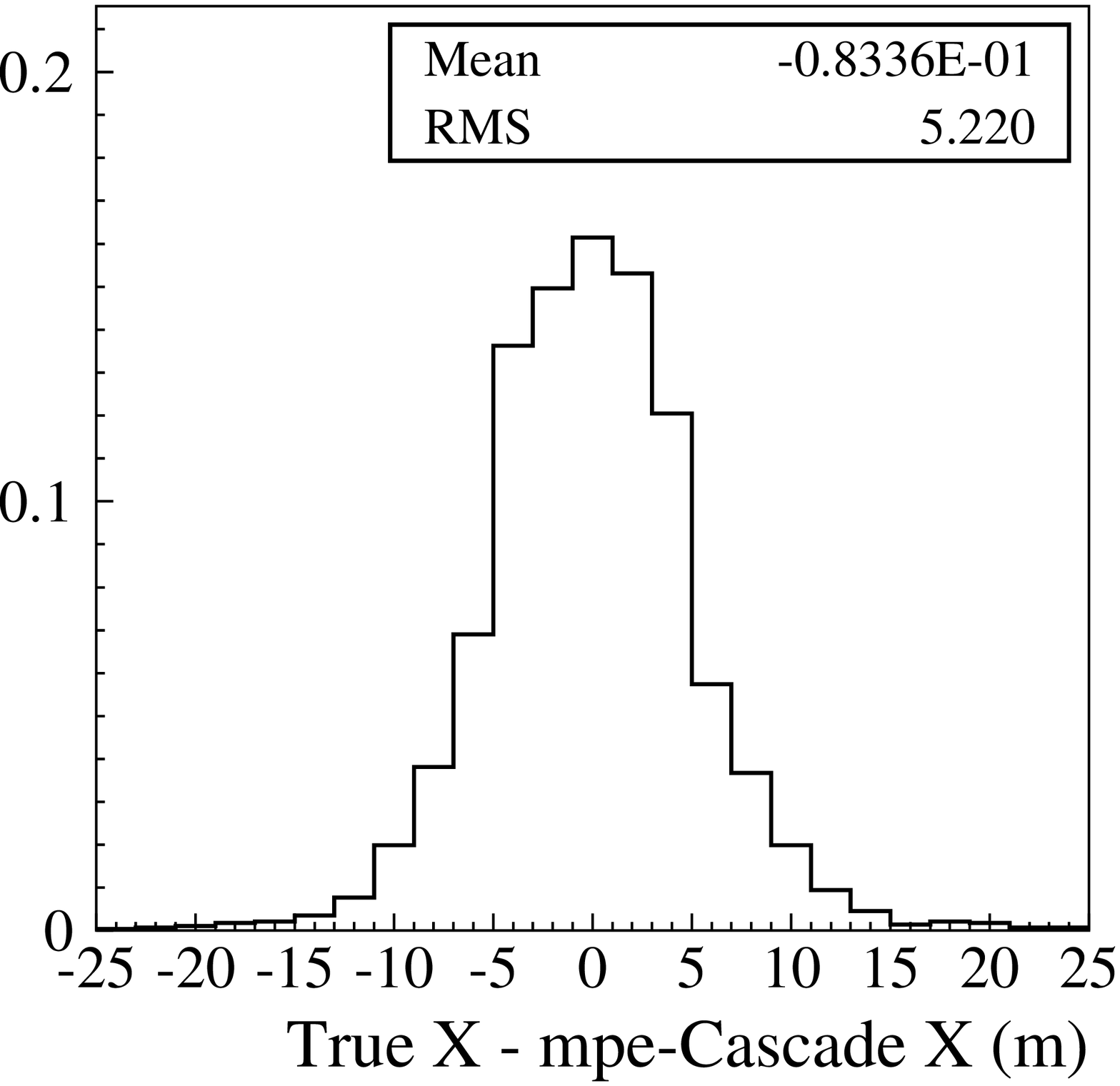}
\includegraphics[width=0.23\textwidth]{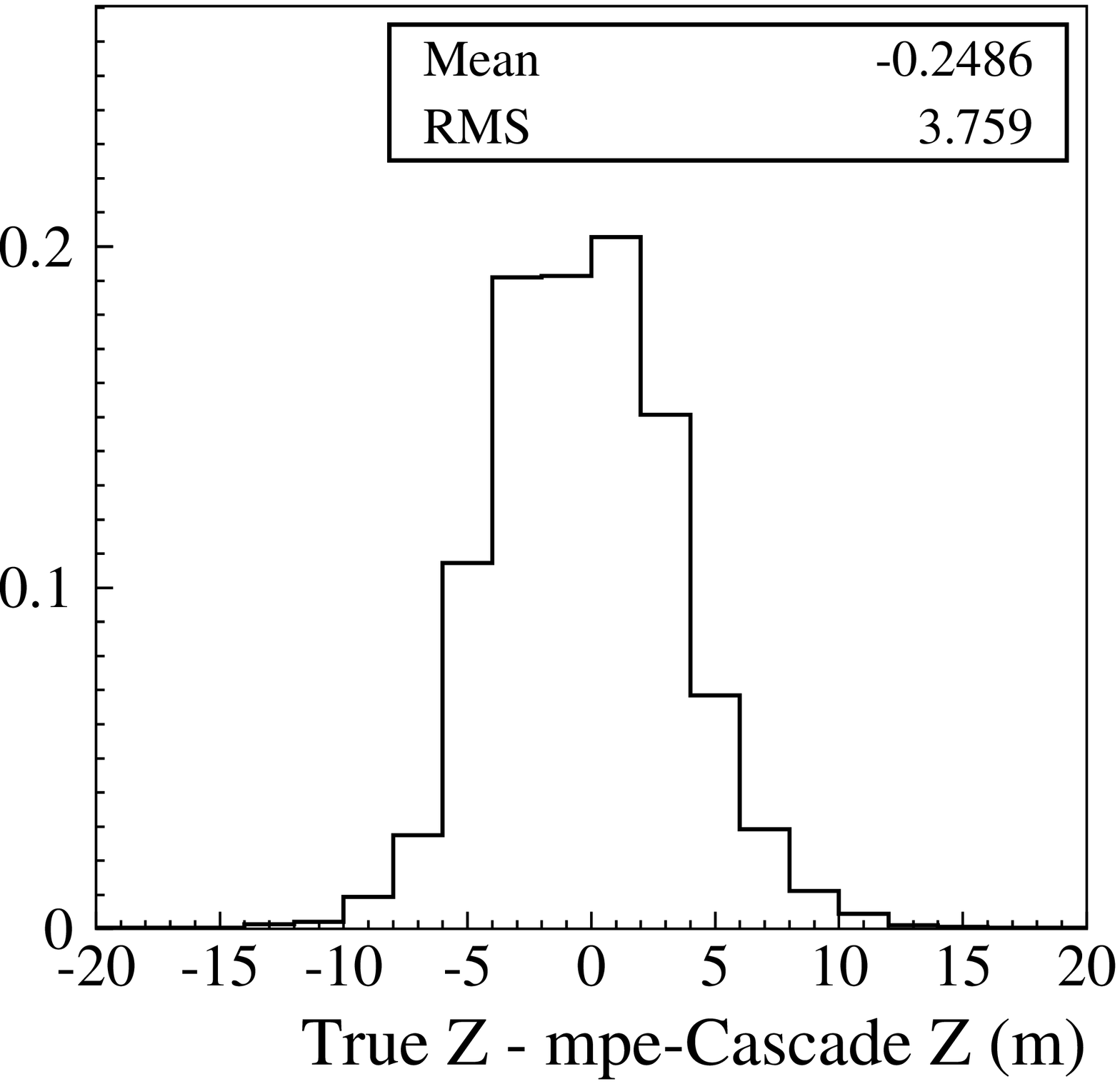}}
\mbox{
\includegraphics[width=0.23\textwidth]{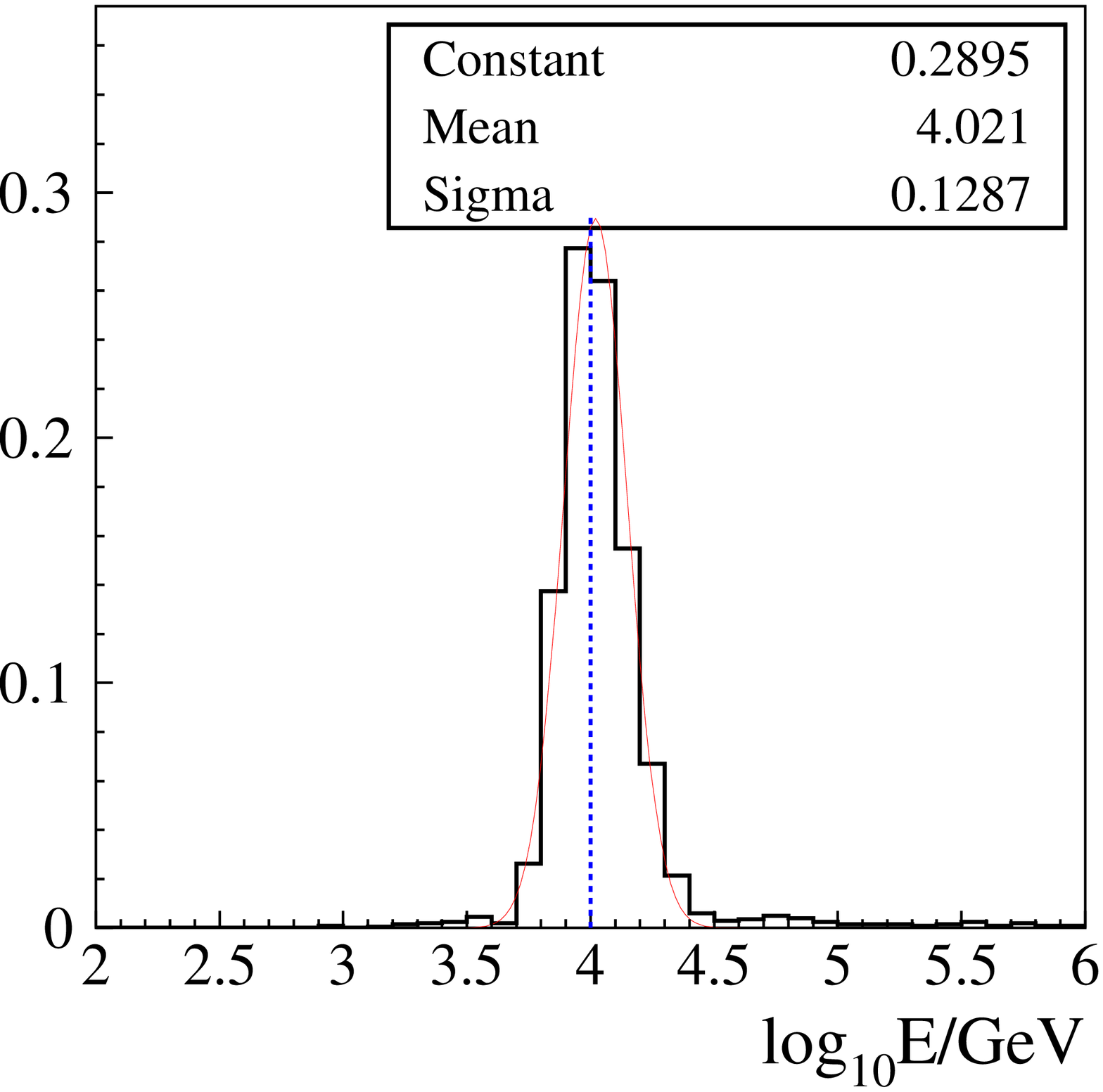}
\includegraphics[width=0.23\textwidth]{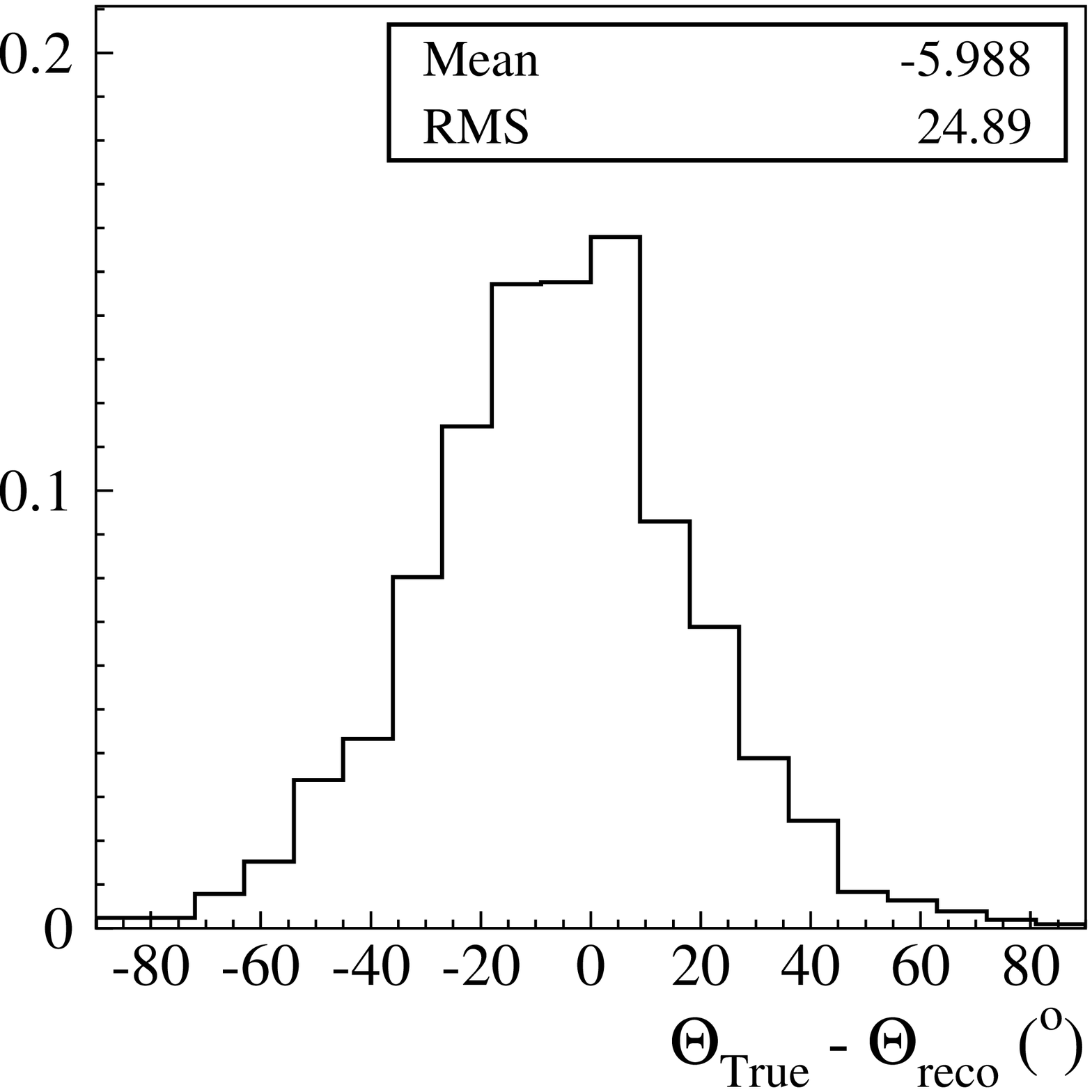}}
\caption{ \label{reco_mc_fig} The four plots show the difference
between simulated and reconstructed vertex position, energy
and direction of cascades. The monoenergetic cascades have $E=10$~TeV
and are contained within the detector. Vertical scale for all four
plots is arbitrary.} 
\end{figure}

Neutrino-induced cascades are reconstructed following the procedure
described in Section~\ref{reco}.  Position, zenith angle and energy
resolutions for a flux of $\nu_e + \overline{\nu}_e$ are calculated
using the difference distributions shown in Fig.~\ref{reco_mc_fig}.
The position resolution is roughly 4~m in the $z$ dimension and 5.5~m
in $x$ and $y$ for contained cascades.  (Note that position resolution
obtained with the pulsed laser is better than that predicted for
Cherenkov light because optical ice properties are more favorable at
the longer wavelength.)  The reconstructed position is biased in the
direction of the cascade, but since the mean of this shift is only
about 2~m for contained cascades it has a negligible impact on the
final result.  Zenith angle resolution is 25$^\circ$--30$^\circ$
depending on the cascade energy.

The energy reconstruction has a resolution of $0.12$--$0.20$ in ${\rm
log}_{10}E$ for contained cascades in the range 1~TeV - 100~TeV,
increasing as a function of cascade energy. Energy reconstruction of
contained cascades is possible from approximately 50~GeV (the minimum
energy cascade which can trigger the detector) to about 100~TeV. At
energies higher than 100~TeV all, or almost all, of the OMs are hit
and thus energy reconstruction by the minimization of
Eq.~\ref{eq:phit} is not possible.  (Such high-energy events would,
however, certainly be identifiable, and probably they can be
reconstructed by other techniques.)

\section{Analysis\label{anal}}

\subsection{\label{filter}Filter}

The first step in the analysis is to apply an initial set of selection
criteria, here called the ``filter,'' which results in a reduction of
the data sample size by more than two orders of magnitude. The filter
first removes spurious hits arising from electronic and PMT noise. It
then uses the fast reconstruction algorithms described earlier, a
simple energy estimator, and topological characteristics of the hit
pattern to select potential signal events.  The various filter steps
were tuned to reject a simulated background of down-going cosmic-ray
muons.

After the filter has been applied, hits likely to have come from
cross-talk are removed, as explained below. Then each event is
reconstructed twice, first with a cascade hypothesis and then with a
muon hypothesis. Several selection criteria are then applied to the
data based on the results of the reconstruction.

Atmospheric $\nu_e$ and $\nu_\mu$ neutrinos are simulated according to
the flux calculated by Lipari~\cite{lipari}. Contributions to cascades
from both atmospheric $\nu_e$ (CC and NC
interactions) and $\nu_\mu$ (NC interactions) are taken
into account. In the energy range relevant to this analysis
($E>$5~TeV), neutrino oscillations are not important in the simulation
of atmospheric neutrinos.

\subsection{\label{xtalk}Removal of Electronic Cross-Talk}

\begin{figure}
\includegraphics[width=0.46\textwidth]{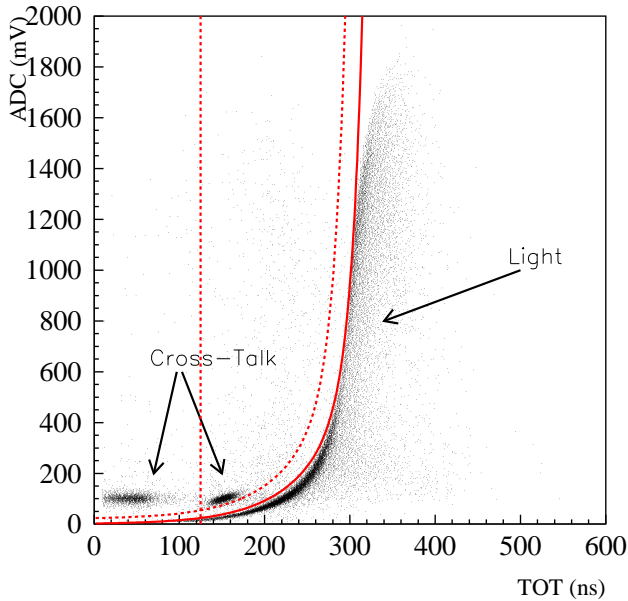}
\caption{\label{fig:xtalk} 
Amplitude vs. time over threshold (TOT) distribution for hits on
OM~149 due to light and cross-talk. The data shown to the right of the
solid curve are generated using the pulsed laser with only the high
voltage for the PMT in OM~149 enabled (all other PMTs had their high
voltage disabled).  This region of the plot therefore contains hits
created by light.  The data shown to the left of the solid curve are
also generated using the pulsed laser, but with a diffuser ball
located 200~m above OM~149, close to OM~129, in the same string as
OM~149. Only OM~129 had its high voltage enabled. This region of the
plot therefore contains hits in OM~149 due to cross-talk.  Hits are
removed from an event if they lie to the left of the dashed curve or
if they have a TOT smaller than 125~ns (indicated by the dashed
line).}
\end{figure}

Electronic cross-talk is present in the twisted pair cables used for
strings 5--10. Spurious hits arising from cross-talk can degrade the
reconstruction quality. Cross-talk is not included in the 
simulations, so its removal is an important facet of this analysis.

We generate a detector-wide map of the cross-talk using the pulsed
laser.  For each OM in strings 5--10, pulsed laser data are taken in
which only the PMT in the OM near the laser diffuser ball has its high
voltage enabled. Any hit in the flashed OM is thus known to be due to
light and any hit in any other OM is known to be due to
cross-talk. The cross-talk map identifies the pairs of
cross-talk-inducing and cross-talk-induced OMs as well as the
correlation in time and amplitude of real and cross-talk hits. The map
shows that cross-talk occurs for OMs in the same string that are close
neighbors or for OMs in the same string that are separated by
150--200~m. The origin of cross-talk is correlated with the relative
positioning of individual electrical cables within the
string~\cite{klug:diplm}.

Cross-talk hits may also be characterized by narrowness (small time over
threshold or ``TOT'') coupled with unexpectedly large
amplitude. Cross-talk-induced and light-induced hits lie in different
regions of the amplitude vs. TOT space and may therefore be separated
from one another. Figure~\ref{fig:xtalk} shows the amplitude vs. TOT
distributions for both types of hits.

The cross-talk map and the amplitude vs. TOT information are both
used to remove cross-talk from the experimental data.

\subsection{Selection Criteria}

\begin{table*}
\footnotesize{
\begin{tabular}{|c|c|c|c|c|c|c|c|}
\hline
   & Selection Criteria & Exp Data & Atm $\mu$ & Atm $\nu_e$ & Atm $\nu_\mu$ &  $\nu_l+\overline{\nu}_l$ & $\nu_e+\overline{\nu}_e$ \\
\hline \hline
   & Trigger                 & 1.05 $\cdot 10^9$ & 1.51 $\cdot 10^8$ & 369.9 & 245.5 & 12446 & 12150 \\ \hline
1  & $N_{\rm big\,TOT} \ge 6$ & & & & & &\\
2  & $v_{\rm line} < 0.12$ or $N_{\rm hits} \ge 75$    & & & & & & \\
3  & $\lambda_1/\lambda_3 > 0.35$ & & & & & & \\
4  & $N_{\rm dir}^{\rm spe} \geq 8$ or $N_{\rm hits} \ge 75$ & & & & & & \\
5  & $L_{\vec{x},t}^{\rm spe} < 7.4$ or $N_{\rm hits} \ge 75$ & 5.57 $\cdot 10^6$ & 1.12$\cdot 10^6$ & 51.1 & 40.4 & 2727 & 3424\\ \hline
6  & $L_{\vec{x},t}^{\rm mpe} < 7.1$ & 1.50 $\cdot 10^6$ &2.03 $\cdot 10^5$ & 38.2 & 30.7 & 2128 & 2818 \\ \hline
7  & $N_{\rm dir}^{\rm mpe} \ge 12 $ & 1.26 $\cdot 10^6$ & 1.14 $\cdot 10^5$ & 32.1& 26.2 & 1909 & 2520 \\ \hline
8  & $\theta_\mu > 80^\circ$ & 3.62 $\cdot 10^5$  & 2.47 $\cdot 10^4$ & 22.5 & 18.5 & 1105 & 1676 \\ \hline
9  & Slices in $z_c$ & 1.48 $\cdot 10^5$  & 1.19 $\cdot 10^4$ & 10.6 & 8.5 & 528 & 711 \\ \hline
10 & cos($\theta_c$) $<$ -0.6 & 675 & 84 & 1.3 & 1.0 & 106 & 156 \\ \hline
11 & $E_c$ vs. $\rho^{\rm mpe}$ & 0 & 0 & 0.01 & 0.01 & 28.7 & 43\\ \hline
\end{tabular}
\caption{ \footnotesize {\label{cut_table}
Selection criteria used in the search for high-energy $\nu$-induced
cascades with AMANDA-B10. The number of events left after applying
each selection criterion to experimental data and background
simulations of atmospheric $\mu$, $\nu_e$ and $\nu_\mu$ are shown. We
simulated 20.3~days of atmospheric $\mu$ data, and 130.1~days of
atmospheric $\nu_e$ and $\nu_\mu$.  Signal simulation is also shown
for $\nu_l+\overline{\nu}_l$ and $\nu_e+\overline{\nu}_e$ assuming
$E^{-2}$ spectra, a flux of $1 \times
10^{-4}~\mathrm{GeV^{-1}\,cm^{-2}\,s^{-1}\,sr^{-1}}$.}  }} 
\end{table*}

Table~\ref{cut_table} lists the selection criteria and the passing rates
for experimental data and the various samples of Monte Carlo used in
this analysis.  Selection criteria which have not already been
described in Section~\ref{reco} are described below, followed by a
physical justification for each criterion.

The ratio of the smallest to the largest eigenvalues of the tensor of
inertia of the position of the hits, $\lambda_1/\lambda_3$ or 
\textit{sphericity}, is used to classify events\footnote{To calculate
a tensor of inertia in this context, a unit mass is hypothesized at
each hit OM position.}. Small values of the sphericity correspond to
hits located along a narrow cylinder, as expected for a muon. Values
of the sphericity close to unity correspond to a spherical
distribution of the hits, as expected for contained cascades.

The dispersive nature of the ice and of the cables that transmit the
electrical signals from the OMs to the electronics on the surface
render sharp signals in the ice into significantly broader pulses at
the surface.  For this reason, counting the number of OMs with large
TOT, $N_{\rm big\,TOT}$, gives a rough estimate of the energy of the
event. For this analysis a TOT is considered large if it exceeds the
value estimated from Monte Carlo simulations to correspond to one
photoelectron by at least a factor of 1.5.  A contained cascade with
300~GeV of energy corresponds roughly to $N_{\rm big\,TOT} = 6$.

The quality of the likelihood reconstruction is determined
from the reduced likelihood parameter, defined by
$L=-log\mathcal{L}/(N_{\rm hits}-N_{\rm fit})$, where $N_{\rm fit}$ is the
number of fitted parameters. Lower values of $L$ correspond to better
reconstruction quality.

A hit is considered \textit{direct} if the time residual is between
-15~ns and 75~ns. The number of direct hits, $N_{\rm dir}$, is another
measure of the quality of the reconstruction. Both the single and
multi-photoelectron position-time reconstruction report the number of
direct hits.

\begin{figure}[t]
\centering
\includegraphics[width=0.46\textwidth]{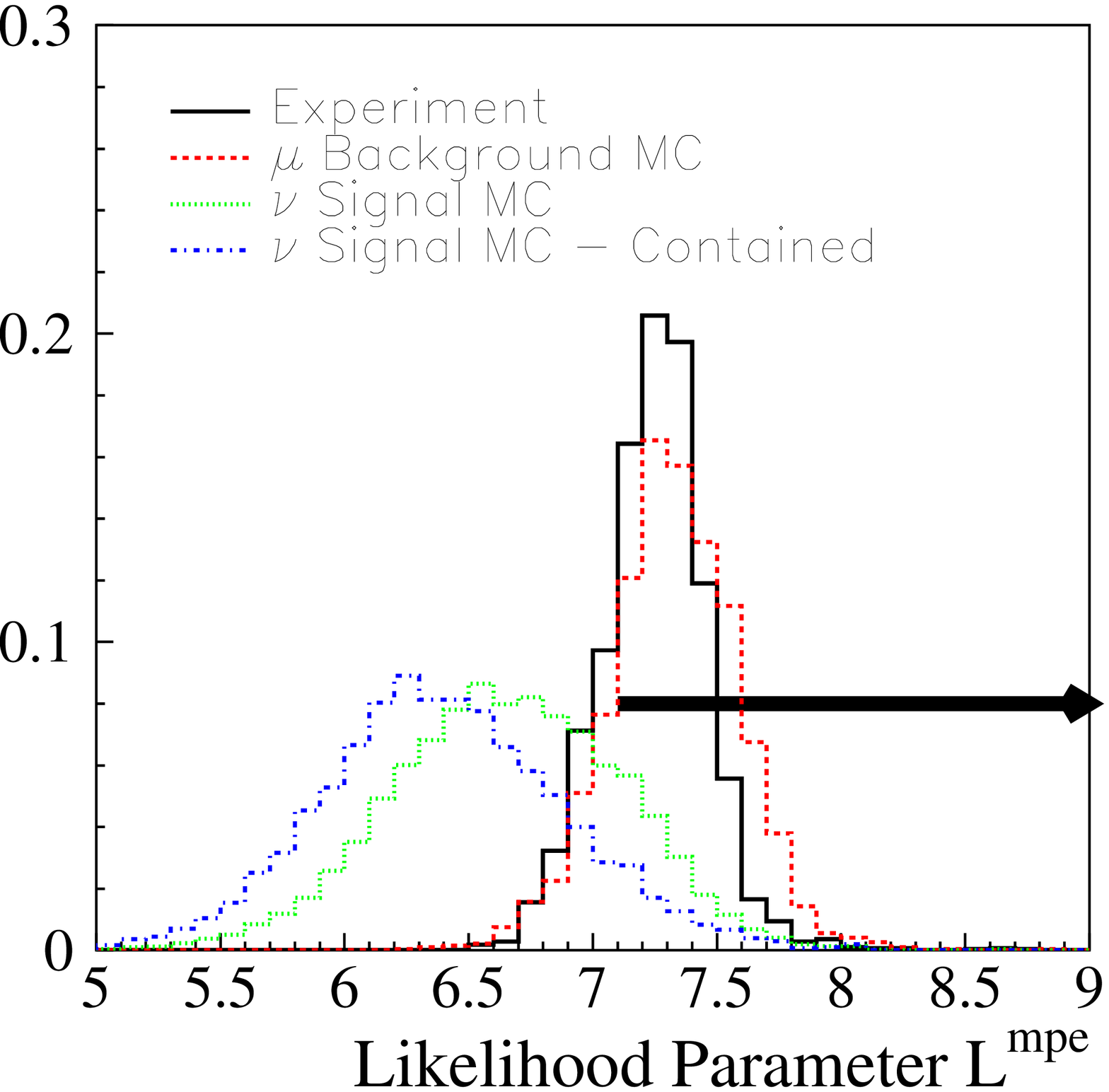}
\caption{\label{fig:ndir-2} 
Normalized distribution of the reduced likelihood parameter $L^{\rm
mpe}_{\vec{x},t}$ for experimental data (solid line), background
atmospheric $\mu$ simulation (dashed line), $\nu_e$  simulation
assuming an $E^{-2}$ power law spectrum (dotted  line) and contained
$\nu_e$  simulation assuming an $E^{-2}$ power law spectrum
(dotted-dashed line). Contained events have their vertices in a
cylinder 400~m in height and 60~m in radius, roughly matching the
detector dimensions. Selection criteria 1-5 from the same table have
already been applied to all the samples shown. The arrow indicates the
region removed by cut 6 from table~\ref{cut_table}.}
\end{figure}

\begin{figure}[t]
\centering
\includegraphics[width=0.46\textwidth]{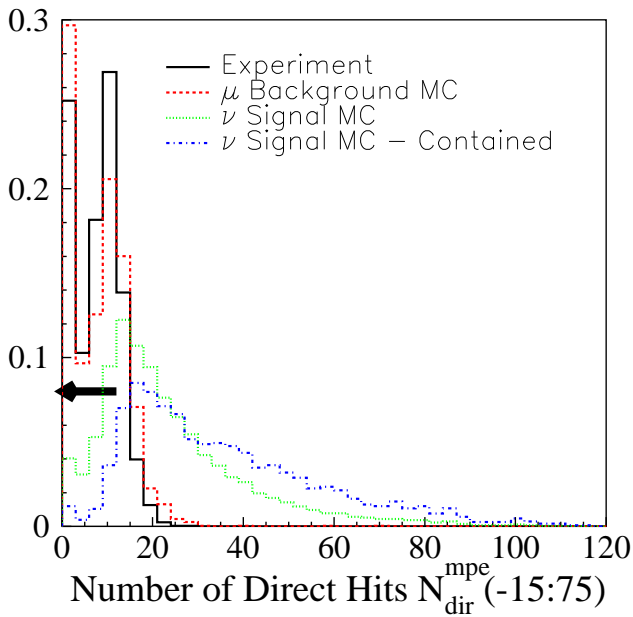}
\caption{\label{fig:ndir} 
Normalized distribution of direct hits for experimental data
(solid line), background atmospheric $\mu$ simulation (dashed line),
$\nu_e$  simulation assuming an $E^{-2}$ power law spectrum (dotted 
line) and contained $\nu_e$  simulation assuming an $E^{-2}$ power law
spectrum (dotted-dashed line). Contained events have their vertices in
a cylinder 400~m in height and 60~m in radius, roughly matching the
detector dimensions. Selection criteria 1-5 from the same table have
already been applied to all the samples shown. The arrow indicates the
region removed by cut 7 from table~\ref{cut_table}.} 
\end{figure}

Criteria 1-5 of table~\ref{cut_table} correspond to the filter and
these criteria must be satisfied by all events in the analysis. The
cut on $N_{\rm big\,TOT}$ selects high-energy events, and the cut on
the sphericity selects events in which the hit topology is
cascade-like. The cut on the speed of the line fit, $v_{line}$,
removes easily identified down-going muons.  As shown in
Fig.~\ref{fig:ndir-2} and Fig.~\ref{fig:ndir}, cuts on the likelihood 
parameter (criterion 6) and the number of direct hits (criterion 7)
are used to eliminate non-cascade-like events and preserve
cascade-like events with good reconstruction quality.  In addition to
the filter criteria, angular cuts (criteria 8 and 10) are used to
reduce clear muon-like events and to select up-going cascade-like
events, and criterion 11 selects high-energy events within a given
distance from the vertical axis of the detector, where $E_c$ is the
reconstructed cascade energy using Eqs.~\ref{eq:mpe}
and~\ref{eq:phit}, and 
$\rho^{\rm mpe} = \sqrt{(x^{\rm mpe})^2 + (y^{\rm mpe})^2 }$.  
(Criterion 9 is discussed in detail below.)

The filter was developed based strictly on the predictions of the
signal and background Monte Carlo. As more and more cuts were applied
after the filter, it was found that the experimental data and
simulations disagreed in the shape of the z-component of the
reconstructed cascade position and in the reconstructed cascade
direction. Inadequately simulated or unsimulated detector
instrumentals, such as optical properties of the ice and cross-talk,
contribute to this disagreement. Restricting the regions of the
detector used in this analysis reduces the effective volume by more
than a factor of two, but restores the agreement between experimental
data and simulations. Moreover, the disagreement is also present in
the reconstructed cascade direction. Only the regions of the detector
that are accepted, show agreement in the reconstructed cascade
direction. Events are accepted only if their reconstructed vertices
satisfy selection criterion number 9: $-80~{\rm m} \leq z_c 
\leq -40~{\rm m}$ or $40~{\rm m} \leq z_c \leq 160~{\rm m}$ with
respect to the center of the detector (located at 1730~m below the
surface).

\section{\label{systematics}Systematic Uncertainties}

There are several uncertainties inherent in estimating the detector sensitivity
to high-energy neutrino-induced cascades. First, the detection medium is 
a natural material (South Pole ice) whose properties are not precisely
known. Second, there are no sufficiently powerful accelerator-based
sources of neutrinos available for use as calibration
beams. Consequently, the understanding of the detector sensitivity is
achieved using down-going atmospheric muons, \textit{in-situ} light
sources and Monte Carlo simulations.

To estimate the systematic uncertainty due to imprecise
knowledge of the optical properties of the ice,
simulations have been performed using the least and the most
transparent ice that we have measured
at AMANDA depths. The cascade sensitivity is modified by 20\% using either
extreme model of the optical properties.  Uncertainties in the bubble
density in the drill-hole ice translate to uncertainties in the OM
angular sensitivity. Monte Carlo simulations with increased bubble
density in the drill-hole ice degrade the cascade sensitivity by 9\%. The
absolute sensitivity of the OMs is also uncertain at the level of
40\%. Monte Carlo simulations with altered absolute OM sensitivity
modifies the cascade sensitivity by 5\%.
This dependence is weak due to the high N$_{\rm hits}$ requirement
imposed by this analysis. (The dependence is much stronger at earlier
stages of the analysis, where the average N$_{\rm hits}$ is much
lower.  For example, before the filter is applied a variation in
absolute OM sensitivity of 40\% results in a modification of the
cascade sensitivity by roughly 35\%.)

Cross-talk can reduce the sensitivity of the detector to high-energy
neutrino-induced cascades. Events for which cross-talk is not fully
removed are typically mis-reconstructed and are therefore unlikely to
have sufficient quality to pass our selection criteria.  The pulsed
laser data is used to estimate the cascade sensitivity loss due to cross-talk
for different locations in the detector. These studies indicate that
the sensitivity is degraded by 7\% due to cross-talk.  (This 7\%
degradation is applied directly to the limit and not treated as a
systematic uncertainty.)  Related to cross-talk is the uncertainty in
the limits due to using slices in $z_c$. Changing each boundary of the
slices by the position resolution in $z$ modifies the cascade sensitivity by
4\%.

Uncertainties in the limits due to neutrino-nucleon cross sections,
total cascade light output, and cascade longitudinal development have
also been estimated using Monte Carlo simulations. For each of
these cases the cascade sensitivity is modified by $< 5$\%. 

The systematic uncertainties discussed so far are added in quadrature,
giving an overall systematic uncertainty on the sensitivity of 25\%.
We follow the procedure described in~\cite{conrad-1,conrad-2} to determine how to
modify the final limit in light of this systematic uncertainty,
assuming that the uncertainties are of a Gaussian nature.

The spectrum of cascade-like events produced by down-going muons is
shown in Fig.~\ref{fig:brem} (see also Section~\ref{brem}). Standard
simulations as well as simulations with modified ice properties and OM
angular and absolute sensitivities have been performed. The
disagreement between experiment and simulations may be explained by
the uncertainties in the knowledge of the optical properties of ice,
the OM sensitivity, the cosmic-ray spectrum and the rate of muon
energy losses.  From Fig.~\ref{fig:brem} it can be seen that
reasonable agreement between experiment and simulations is restored by
shifting the energy scale by up to 0.2 in $\log_{10}E$. This
uncertainty in the energy scale results in an uncertainty on the
sensitivity of less than 25\%. This uncertainty is \textit{not} independent of
the other sources of systematic uncertainty that we have studied.  It
demonstrates, however, that the overall systematic uncertainty has not
been grossly under- or overestimated.

\section{Results}

The analysis is applied to simulated samples of atmospheric $\nu_e$
and $\nu_\mu$ background, high-energy neutrino signal (all flavors),
and atmospheric muons, and to the 1997 experimental data set. In the
experimental data zero events are found. The simulation of atmospheric
$\nu_e$ predicts 0.01 events, and the simulation of atmospheric
$\nu_\mu$ predicts 0.01 events from NC interactions (both these
numbers have been rounded up from distinct smaller values). Zero
events are found in the simulated atmospheric muon sample after all
cuts. A limit on the flux of neutrinos assuming an $E^{-2}$ power law
spectrum is set using the following formula:

\begin{widetext}
\begin{equation}
\label{limit_eq}
E^2 \frac{d\Phi}{dE} = \frac{N_{\rm 90\%}}{T N_{A} \rho_{\rm ice}
\sum_l f_l \int E^{-2} \xi_l(E,\theta) \sigma^l_{\rm tot}(E) V^l_{\rm
eff}(E,\theta) \, d\Omega dE} 
\end{equation}
\end{widetext}

\noindent where $l$ is the neutrino flavor, $E$ the neutrino energy, 
$\theta$ the neutrino zenith angle, $N_{\rm 90\%}=2.62$ determined
using the unified Feldman-Cousins procedure~\cite{feldman} with a
correction applied for the estimated 25\% systematic
uncertainty~\cite{conrad-1,conrad-2}, $T$ the live-time (130.1 days), $N_{A}$
Avogadro's number, $\rho_{\rm ice}$ the density of ice, $\sigma^l_{\rm
tot}(E)$ the neutrino cross section~\cite{gandhi}, $V^l_{\rm
eff}(E,\theta)$ the effective volume of the detector (see
table~\ref{table:effec_vol}), $f_l$ the fraction of the total neutrino
flux comprised by the neutrino flavor $l$, and $\xi_l(E,\theta)$ a
function that corrects the flux for Earth absorption and NC
scattering. The integration of Eq.~\ref{limit_eq} has been done for
neutrino energies between 5~TeV and 300~TeV.

The 90\% C.L. limit on the diffuse flux of $\nu_e+\nu_\mu+\nu_\tau+
\overline{\nu}_e+\overline{\nu}_\mu+\overline{\nu}_\tau$ for
neutrino energies between 5~TeV and 300~TeV, assuming a neutrino flux
ratio of 1:1:1 at the detector, is:

\begin{equation}
   \label{nu_l_limit_eq}
   E^2\frac{d\Phi}{dE} < 9.8 \times 10^{-6} \; \mathrm{GeV\,cm^{-2}\,s^{-1}\,sr^{-1}}.
\end{equation}

\noindent The 90\% C.L. limit on the diffuse flux of $\nu_e+\overline{\nu}_e$ for
neutrino energies between 5~TeV and 300~TeV
is:
\begin{equation}
   \label{nu_e_limit_eq}
   E^2\frac{d\Phi}{dE} < 6.5 \times 10^{-6} \;
   \mathrm{GeV\,cm^{-2}\,s^{-1}\,sr^{-1}}. 
\end{equation}
The latter limit is independent of the assumed neutrino flux ratio.
The limits without incorporating the effects of systematic
uncertainties are $9.1 \times
10^{-6}~\mathrm{GeV\,cm^{-2}\,s^{-1}\,sr^{-1}} $ and $6.1 \times
10^{-6}~\mathrm{GeV\,cm^{-2}\,s^{-1}\,sr^{-1}} $, respectively.  (Note
that since the limit in Eq.~\ref{nu_l_limit_eq} is on the sum of the
fluxes of all neutrino flavors, and the limit in
Eq.~\ref{nu_e_limit_eq} is on an individual flavor, the former limit
should be divided by a factor of three to compare it properly to the
latter.)

\begin{figure}
\includegraphics[width=0.46\textwidth]{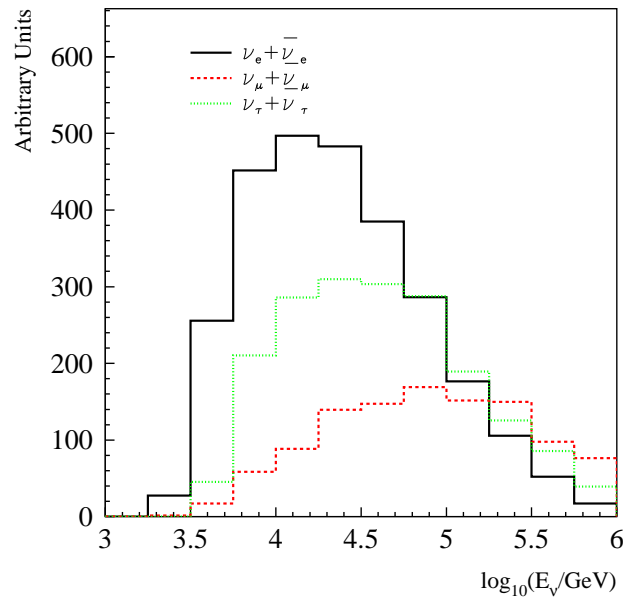}
\caption{\label{fig:final_energy} 
Distribution of $\nu_e$, $\nu_\mu$ and $\nu_\tau$ energies after all
selection criteria have been applied.  The relative normalization
between the histograms indicates the relative number of events for
each neutrino flavor that passes all the selection criteria.  The initial
energy distribution follows an $E^{-2}$ spectrum. Neutrino absorption
inside Earth, NC scattering and $\tau$ decay have been
taken into account as described in Section~\ref{mc_pred}.}
\end{figure}

\begin{figure}
\includegraphics[width=0.46\textwidth]{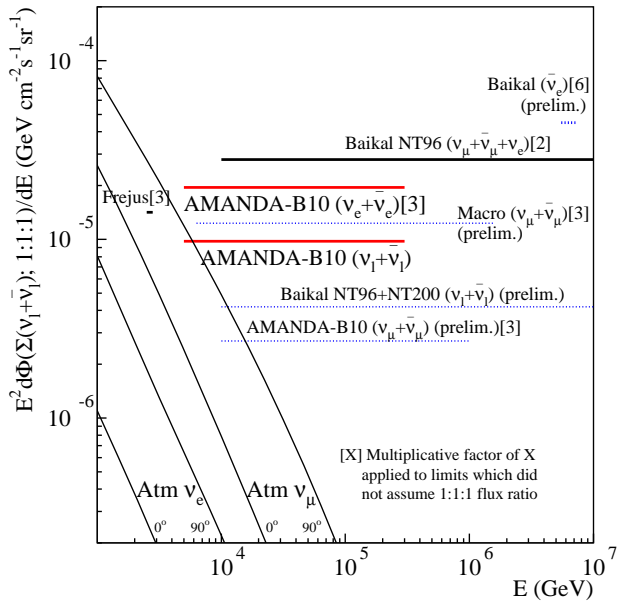}
\caption{\label{fig:limits} 
The limits on the cascade-producing neutrino flux, summed over the
three active flavors, presented in this work and in other experiments,
with multiplicative factors applied as indicated to permit comparison
of limits derived with different assumed neutrino fluxes at the
detector: Baikal ($\overline{\nu}_e$)~\cite{venice01:baikal} (at the
$W^\pm$ resonance); Baikal NT96
($\nu_\mu+\overline{\nu}_\mu+\nu_e$)~\cite{baikal:astropart}; Frejus
($\nu_\mu+\overline{\nu}_\mu$)~\cite{frejus}; MACRO
($\nu_\mu+\overline{\nu}_\mu$)~\cite{icrc01:macro}.  Baikal NT96+NT200
($\nu_l+\overline{\nu}_l$)~\cite{venice01:baikal,jan:baikal}; AMANDA B-10
($\nu_\mu+\overline{\nu}_\mu$)~\cite{icrc01:dif}; Also shown are the
predicted horizontal and vertical $\nu_e$ and $\nu_\mu$ atmospheric
fluxes~\cite{lipari}.}
\end{figure}

Our results together with other limits on the flux of diffuse
neutrinos are shown in Fig.~\ref{fig:limits}.  Since recent results
from other low energy neutrino
experiments~\cite{sno-cc,sno-nc,sno-dn,superk-atm} indicate that
high-energy cosmological neutrinos will have a neutrino flavor flux
ratio of 1:1:1 upon detection, in this figure we scale limits derived
under different assumptions accordingly.  For example, to do a
side-by-side comparison of a limit on the flux of
$\nu_e+\nu_\mu+\nu_\tau+\overline{\nu}_e+\overline{\nu}_\mu+\overline{\nu}_\tau$,
derived under the assumption of a ratio of 1:1:1, to a limit on just
the flux of $\nu_\mu+\overline{\nu}_\mu$, the latter must be degraded
by a factor of three.  (N.B.: We assume that
$\nu$:$\overline{\nu}$::1:1.) Following the Learned and Mannheim
prescription for presenting limits~\cite{learned-review}, we show
neutrino energy distributions after applying all the selection
criteria in Fig.~\ref{fig:final_energy}.

It should be noted that most searches of diffuse fluxes shown in
Fig.~\ref{fig:limits} are based on the observation of up-going
neutrino-induced muons. Only Baikal and AMANDA have presented limits
from analyses that search for neutrino-induced cascades and only the
AMANDA analysis uses full cascade event reconstruction.

\section{Conclusions}

High-energy neutrino-induced cascades have been searched for in the
data collected by AMANDA-B10 in 1997. Detailed event reconstruction
was performed.  Using \textit{in-situ} light sources and atmospheric muon
catastrophic energy losses, the sensitivity of the detector to
high-energy cascades has been demonstrated.

No evidence for the existence of a diffuse flux of neutrinos producing
cascade signatures has been found. Effective volumes as a function of
energy and zenith angle for all neutrino flavors have been
presented. The effective volumes allow the calculation of limits for
any predicted neutrino flux model. The limit on cascades from a
diffuse flux of
$\nu_e+\nu_\mu+\nu_\tau+\overline{\nu}_e+\overline{\nu}_\mu+\overline{\nu}_\tau$
is $E^2 \frac{d\Phi}{dE} < 9.8 \times
10^{-6}\;\mathrm{GeV\,cm^{-2}\,s^{-1}\,sr^{-1}}$ assuming a neutrino
flavor flux ratio of 1:1:1 at the detector.  The limit on cascades
from a diffuse flux of $\nu_e+\overline{\nu}_e$ is $E^2 \frac{d\Phi}{dE} < 6.5
\times 10^{-6}\;\mathrm{ GeV\,cm^{-2}\,s^{-1}\,sr^{-1}}$, independent 
of the assumed neutrino flux ratio.  The limits are valid for
neutrino fluxes in the energy range of 5~TeV to 300~TeV.

\begin{table*}
\footnotesize{
\begin{center}
\begin{tabular}{|c|c|c|c|c|c|}
\hline
        &                        & 3.0-10.0~TeV  & 10.0-30~TeV   & 30-100~TeV    & 100-300~TeV\\
\hline \hline
        & $-1<\cos\theta<-0.6$   & $0.80\pm0.05$ & $1.85\pm0.10$ & $1.87\pm0.15$ & $1.37\pm0.20$ \\
$\nu_e$ & $-0.6<\cos\theta<-0.2$ & $0.40\pm0.03$ & $0.85\pm0.07$ & $1.10\pm0.10$ & $0.72\pm0.10$ \\
        & $-0.2<\cos\theta<0.2$  & $0.08\pm0.01$ & $0.22\pm0.02$ & $0.36\pm0.05$ & $0.31\pm0.07$ \\
\hline
                   & $-1<\cos\theta<-0.6$   & $0.82\pm0.05$ & $1.67\pm0.12$  & $1.85\pm0.10$ & $1.60\pm0.15$ \\
$\overline{\nu}_e$ & $-0.6<\cos\theta<-0.2$ & $0.42\pm0.03$ & $0.77\pm 0.07$ & $0.92\pm0.07$ & $0.74\pm0.10$   \\
                   & $-0.2<\cos\theta<0.2$  & $0.09\pm0.01$ & $0.20\pm0.02$  & $0.35\pm0.05$ & $0.30\pm0.07$\\
\hline
          & $-1<\cos\theta<-0.6$   & $0.08\pm0.02$ & $0.35\pm0.05$ & $0.87\pm0.1$  & $1.27\pm0.15$ \\
$\nu_\mu$ & $-0.6<\cos\theta<-0.2$ & $0.05\pm0.01$ & $0.25\pm0.03$ & $0.70\pm0.10$ & $1.60\pm0.10$ \\
          & $-0.2<\cos\theta<0.2$  & $-$           & $-$           & $-$           & $0.05\pm0.01$\\
\hline
                     & $-1<\cos\theta<-0.6$   & $0.12\pm0.02$ & $0.34\pm0.05$ & $0.70\pm0.05$ & $1.17\pm0.15$\\
$\overline{\nu}_\mu$ & $-0.6<\cos\theta<-0.2$ & $0.05\pm0.01$ & $0.25\pm0.03$ & $0.70\pm0.1$  & $0.14\pm0.01$ \\
                     & $-0.2<\cos\theta<0.2$  & $-$           & $-$           & $-$           & $0.03\pm0.01$\\
\hline
           & $-1<\cos\theta<-0.6$   & $0.35\pm0.05$ & $1.10\pm0.10$ & $1.85\pm0.15$ & $1.35\pm0.20$ \\
$\nu_\tau$ & $-0.6<\cos\theta<-0.2$ & $0.15\pm0.03$ & $0.50\pm0.05$ & $0.85\pm0.10$ & $1.05\pm0.10$\\
           & $-0.2<\cos\theta<0.2$  & $0.04\pm0.01$ & $0.10\pm0.02$ & $0.23\pm0.05$ & $0.32\pm0.07$\\
\hline
                      & $-1<\cos\theta<-0.6$   & $0.35\pm0.05$ & $1.15\pm0.10$ & $1.65\pm0.10$ & $1.50\pm0.15$\\
$\overline{\nu}_\tau$ & $-0.6<\cos\theta<-0.2$ & $0.15\pm0.03$ & $0.45\pm0.05$ & $0.80\pm0.10$ & $1.20\pm0.10$ \\
                      & $-0.2<\cos\theta<0.2$  & $0.06\pm0.01$ & $0.12\pm0.02$ & $0.22\pm0.04$ & $0.31\pm0.06$\\
\hline
\end{tabular}
\end{center}
\caption{\label{table:effec_vol}
Effective volume, in units of $10^{-3}$km$^3$, for all neutrino flavors
as a function of energy and zenith angle after all the selection
criteria have been applied. Uncertainties are statistical only.}} 
\end{table*}

\begin{acknowledgments}
This research was supported by the following agencies: U.S.  National
Science Foundation, Office of Polar Programs; U.S. National Science
Foundation, Physics Division; University of Wisconsin Alumni Research
Foundation; U.S. Department of Energy; Swedish Natural Science
Research Council; Swedish Research Council; Swedish Polar Research
Secretariat; Knut and Alice Wallenberg Foundation, Sweden; German
Ministry for Education and Research; U.S.  National Energy Research
Scientific Computing Center (supported by the Office of Energy
Research of the U.S.  Department of Energy); UC-Irvine AENEAS
Supercomputer Facility; Deutsche Forschungsgemeinschaft
(DFG). D.F. Cowen acknowledges the support of the NSF CAREER program
and C. P\'erez de los Heros acknowledges support from the EU 4th
framework of Training and Mobility of Researchers.
\end{acknowledgments}

\bibliographystyle{apsrev}

\begin{thebibliography}{42}
\expandafter\ifx\csname natexlab\endcsname\relax\def\natexlab#1{#1}\fi
\expandafter\ifx\csname bibnamefont\endcsname\relax
  \def\bibnamefont#1{#1}\fi
\expandafter\ifx\csname bibfnamefont\endcsname\relax
  \def\bibfnamefont#1{#1}\fi
\expandafter\ifx\csname citenamefont\endcsname\relax
  \def\citenamefont#1{#1}\fi
\expandafter\ifx\csname url\endcsname\relax
  \def\url#1{\texttt{#1}}\fi
\expandafter\ifx\csname urlprefix\endcsname\relax\def\urlprefix{URL }\fi
\providecommand{\bibinfo}[2]{#2}
\providecommand{\eprint}[2][]{\url{#2}}

\bibitem[{\citenamefont{Ahmad et~al.}(2001)}]{sno-cc}
\bibinfo{author}{\bibfnamefont{Q.}~\bibnamefont{Ahmad}} \bibnamefont{et~al.},
  \bibinfo{journal}{Phys. Rev. Lett.} \textbf{\bibinfo{volume}{87}},
  \bibinfo{pages}{071301} (\bibinfo{year}{2001}).

\bibitem[{\citenamefont{Ahmad et~al.}(2002{\natexlab{a}})}]{sno-nc}
\bibinfo{author}{\bibfnamefont{Q.}~\bibnamefont{Ahmad}} \bibnamefont{et~al.},
  \bibinfo{journal}{Submitted to Phys. Rev. Lett.}
  (\bibinfo{year}{2002}{\natexlab{a}}), \eprint{arXiv:nucl-ex/0204008}.

\bibitem[{\citenamefont{Ahmad et~al.}(2002{\natexlab{b}})}]{sno-dn}
\bibinfo{author}{\bibfnamefont{Q.}~\bibnamefont{Ahmad}} \bibnamefont{et~al.},
  \bibinfo{journal}{Submitted to Phys. Rev. Lett.}
  (\bibinfo{year}{2002}{\natexlab{b}}), \eprint{arXiv:nucl-ex/0204009}.

\bibitem[{\citenamefont{Fukuda et~al.}(2000)}]{superk-atm}
\bibinfo{author}{\bibfnamefont{S.}~\bibnamefont{Fukuda}} \bibnamefont{et~al.},
  \bibinfo{journal}{Phys. Rev. Lett.} \textbf{\bibinfo{volume}{85}},
  \bibinfo{pages}{3999} (\bibinfo{year}{2000}).

\bibitem[{\citenamefont{Andr\'{e}s et~al.}(2001)}]{ama:nature-pub}
\bibinfo{author}{\bibfnamefont{E.}~\bibnamefont{Andr\'{e}s}}
  \bibnamefont{et~al.}, \bibinfo{journal}{Nature}
  \textbf{\bibinfo{volume}{410}}, \bibinfo{pages}{441} (\bibinfo{year}{2001}).

\bibitem[{\citenamefont{Ahrens et~al.}(2002{\natexlab{a}})}]{amanda:atm-b10}
\bibinfo{author}{\bibfnamefont{J.}~\bibnamefont{Ahrens}} \bibnamefont{et~al.},
  \bibinfo{journal}{Accepted for publication by Phys. Rev. D.}
  (\bibinfo{year}{2002}{\natexlab{a}}), \eprint{arXiv:astro-ph/0205109}.

\bibitem[{\citenamefont{Ahrens et~al.}(2002{\natexlab{b}})}]{amanda:wimps-b10}
\bibinfo{author}{\bibfnamefont{J.}~\bibnamefont{Ahrens}} \bibnamefont{et~al.},
  \bibinfo{journal}{Accepted for publication by Phys. Rev. D.}
  (\bibinfo{year}{2002}{\natexlab{b}}), \eprint{arXiv:astro-ph/0202370}.

\bibitem[{\citenamefont{Halzen and Saltzberg}(1998)}]{halzen-saltzberg}
\bibinfo{author}{\bibfnamefont{F.}~\bibnamefont{Halzen}} \bibnamefont{and}
  \bibinfo{author}{\bibfnamefont{D.}~\bibnamefont{Saltzberg}},
  \bibinfo{journal}{Phys. Rev. Lett.} \textbf{\bibinfo{volume}{81}},
  \bibinfo{pages}{4305} (\bibinfo{year}{1998}).

\bibitem[{\citenamefont{Dutta et~al.}(2001)}]{reno}
\bibinfo{author}{\bibfnamefont{S.}~\bibnamefont{Dutta}} \bibnamefont{et~al.},
  \bibinfo{journal}{Phys. Rev. D} \textbf{\bibinfo{volume}{64}},
  \bibinfo{pages}{113015} (\bibinfo{year}{2001}).

\bibitem[{\citenamefont{Beacom et~al.}(2001)}]{beacom}
\bibinfo{author}{\bibfnamefont{J.}~\bibnamefont{Beacom}} \bibnamefont{et~al.}
  (\bibinfo{year}{2001}), \eprint{arXiv:astro-ph/0111482}.

\bibitem[{\citenamefont{Stanev}(1999)}]{stanev:prl}
\bibinfo{author}{\bibfnamefont{T.}~\bibnamefont{Stanev}},
  \bibinfo{journal}{Phys. Rev. Lett.} \textbf{\bibinfo{volume}{83}},
  \bibinfo{pages}{5427} (\bibinfo{year}{1999}).

\bibitem[{\citenamefont{Wiebusch}(1995)}]{wiebusch:phd}
\bibinfo{author}{\bibfnamefont{C.}~\bibnamefont{Wiebusch}}, Ph.D. thesis,
  \bibinfo{school}{RWTH Aachen}, \bibinfo{address}{Aachen, Germany}
  (\bibinfo{year}{1995}).

\bibitem[{\citenamefont{Hill and Leuthold}(2001)}]{icrc01:dif}
\bibinfo{author}{\bibfnamefont{G.}~\bibnamefont{Hill}} \bibnamefont{and}
  \bibinfo{author}{\bibfnamefont{M.}~\bibnamefont{Leuthold}}, in
  \emph{\bibinfo{booktitle}{Proc. 27$^{th}$ Int. Cosmic Ray Conf.}}
  (\bibinfo{address}{Hamburg, Germany}, \bibinfo{year}{2001}), p.
  \bibinfo{pages}{1113}.

\bibitem[{\citenamefont{Hill}(1999)}]{icrc99:b10}
\bibinfo{author}{\bibfnamefont{G.}~\bibnamefont{Hill}}, in
  \emph{\bibinfo{booktitle}{Proc. 26$^{th}$ Int. Cosmic Ray Conf.}}
  (\bibinfo{address}{Salt Lake City, USA}, \bibinfo{year}{1999}), HE.6.3.02.

\bibitem[{\citenamefont{Price et~al.}(2000)}]{price:geophys}
\bibinfo{author}{\bibfnamefont{B.}~\bibnamefont{Price}} \bibnamefont{et~al.},
  \bibinfo{journal}{Geophys. Rev. Lett.} \textbf{\bibinfo{volume}{27}},
  \bibinfo{pages}{2129} (\bibinfo{year}{2000}).

\bibitem[{\citenamefont{Stenger}(1990)}]{stenger}
\bibinfo{author}{\bibfnamefont{V.}~\bibnamefont{Stenger}},
  \bibinfo{type}{DUMAND Internal Report} \bibinfo{number}{HDC-1-90}
  (\bibinfo{year}{1990}).

\bibitem[{\citenamefont{Wiebusch}(1998)}]{muon-reco}
\bibinfo{author}{\bibfnamefont{C.}~\bibnamefont{Wiebusch}}
  (\bibinfo{address}{DESY-Zeuthen, Germany}, \bibinfo{year}{1998}),
  DESY-proc-1999-01.

\bibitem[{\citenamefont{Kowalski and Taboada}(2001)}]{picrc}
\bibinfo{author}{\bibfnamefont{M.}~\bibnamefont{Kowalski}} \bibnamefont{and}
  \bibinfo{author}{\bibfnamefont{I.}~\bibnamefont{Taboada}}, in
  \emph{\bibinfo{booktitle}{Proc. of 2$^{nd}$ Workshop Methodical Aspects of
  Underwater/Ice Neutrino Telescopes}} (\bibinfo{address}{Hamburg, Germany},
  \bibinfo{year}{2001}).

\bibitem[{\citenamefont{Kowalski}(2000)}]{kowalksi:diplm}
\bibinfo{author}{\bibfnamefont{M.}~\bibnamefont{Kowalski}},
  \bibinfo{type}{Diploma thesis}, \bibinfo{school}{Humboldt University},
  \bibinfo{address}{Berlin, Germany} (\bibinfo{year}{2000}).

\bibitem[{\citenamefont{Taboada}(2002)}]{taboada:phd}
\bibinfo{author}{\bibfnamefont{I.}~\bibnamefont{Taboada}}, Ph.D. thesis,
  \bibinfo{school}{University of Pennsylvania}, \bibinfo{address}{Philadelphia,
  USA} (\bibinfo{year}{2002}).

\bibitem[{\citenamefont{Andr\'{e}s et~al.}(2000)}]{ama:b4-pub}
\bibinfo{author}{\bibfnamefont{E.}~\bibnamefont{Andr\'{e}s}}
  \bibnamefont{et~al.}, \bibinfo{journal}{Astro. Part. Phys.}
  \textbf{\bibinfo{volume}{13}}, \bibinfo{pages}{1} (\bibinfo{year}{2000}).

\bibitem[{\citenamefont{Heck et~al.}(1998)}]{corsika}
\bibinfo{author}{\bibfnamefont{D.}~\bibnamefont{Heck}} \bibnamefont{et~al.},
  \bibinfo{type}{Tech. Rep.} \bibinfo{number}{FZKA 6019},
  \bibinfo{institution}{Forshungszebtrum Karlsruhe},
  \bibinfo{address}{Karlsruhe, Germany} (\bibinfo{year}{1998}).

\bibitem[{\citenamefont{Kalmykov et~al.}(1997)}]{Kalmykov}
\bibinfo{author}{\bibfnamefont{N.}~\bibnamefont{Kalmykov}}
  \bibnamefont{et~al.}, \bibinfo{journal}{Nucl. Phys. B (Proc. Suppl.)}
  \textbf{\bibinfo{volume}{52B}}, \bibinfo{pages}{17} (\bibinfo{year}{1997}).

\bibitem[{\citenamefont{Wiebel-Sooth et~al.}(1997)}]{cr-composition}
\bibinfo{author}{\bibfnamefont{B.}~\bibnamefont{Wiebel-Sooth}}
  \bibnamefont{et~al.} (\bibinfo{year}{1997}), \eprint{arXiv:astro-ph/9709253}.

\bibitem[{\citenamefont{Lohmann et~al.}(1985)}]{lohmann-85}
\bibinfo{author}{\bibfnamefont{W.}~\bibnamefont{Lohmann}} \bibnamefont{et~al.},
  \bibinfo{journal}{CERN Yellow Report CERN-85-03}  (\bibinfo{year}{1985}).

\bibitem[{\citenamefont{Kopp et~al.}(1995)}]{lohmann-95}
\bibinfo{author}{\bibfnamefont{R.}~\bibnamefont{Kopp}} \bibnamefont{et~al.},
  \bibinfo{journal}{private comunication}  (\bibinfo{year}{1995}).

\bibitem[{\citenamefont{Dziewonski and Anderson}(1981)}]{earth}
\bibinfo{author}{\bibfnamefont{A.}~\bibnamefont{Dziewonski}} \bibnamefont{and}
  \bibinfo{author}{\bibnamefont{Anderson}}, \bibinfo{journal}{Phys. Earth
  Planet Int.} \textbf{\bibinfo{volume}{25}}, \bibinfo{pages}{297}
  (\bibinfo{year}{1981}).

\bibitem[{\citenamefont{Gandhi et~al.}(1999)}]{gandhi}
\bibinfo{author}{\bibfnamefont{R.}~\bibnamefont{Gandhi}} \bibnamefont{et~al.},
  \bibinfo{journal}{Phys. Rev. D} \textbf{\bibinfo{volume}{58}},
  \bibinfo{pages}{93009} (\bibinfo{year}{1999}).

\bibitem[{\citenamefont{Jadach et~al.}(1990)\citenamefont{Jadach, Kuhn, and
  Was}}]{tauola1}
\bibinfo{author}{\bibfnamefont{S.}~\bibnamefont{Jadach}},
  \bibinfo{author}{\bibfnamefont{J.}~\bibnamefont{Kuhn}}, \bibnamefont{and}
  \bibinfo{author}{\bibfnamefont{Z.}~\bibnamefont{Was}},
  \bibinfo{journal}{Comput. Phys. Commun.} \textbf{\bibinfo{volume}{64}},
  \bibinfo{pages}{275} (\bibinfo{year}{1990}).

\bibitem[{\citenamefont{Jezabek et~al.}(1992)\citenamefont{Jezabek, Was,
  Jadach, and Kuhn}}]{tauola2}
\bibinfo{author}{\bibfnamefont{M.}~\bibnamefont{Jezabek}},
  \bibinfo{author}{\bibfnamefont{Z.}~\bibnamefont{Was}},
  \bibinfo{author}{\bibfnamefont{S.}~\bibnamefont{Jadach}}, \bibnamefont{and}
  \bibinfo{author}{\bibfnamefont{J.}~\bibnamefont{Kuhn}},
  \bibinfo{journal}{Comput. Phys. Commun.} \textbf{\bibinfo{volume}{70}},
  \bibinfo{pages}{69} (\bibinfo{year}{1992}).

\bibitem[{\citenamefont{Jadach et~al.}(1993)\citenamefont{Jadach, Was, Decker,
  and Kuhn}}]{tauola3}
\bibinfo{author}{\bibfnamefont{S.}~\bibnamefont{Jadach}},
  \bibinfo{author}{\bibfnamefont{Z.}~\bibnamefont{Was}},
  \bibinfo{author}{\bibfnamefont{R.}~\bibnamefont{Decker}}, \bibnamefont{and}
  \bibinfo{author}{\bibfnamefont{J.}~\bibnamefont{Kuhn}},
  \bibinfo{journal}{Comput. Phys. Commun.} \textbf{\bibinfo{volume}{76}},
  \bibinfo{pages}{361} (\bibinfo{year}{1993}).

\bibitem[{\citenamefont{Lipari}(1993)}]{lipari}
\bibinfo{author}{\bibfnamefont{P.}~\bibnamefont{Lipari}},
  \bibinfo{journal}{Astro. Part. Phys.} \textbf{\bibinfo{volume}{1}},
  \bibinfo{pages}{193} (\bibinfo{year}{1993}).

\bibitem[{\citenamefont{Klug}(1997)}]{klug:diplm}
\bibinfo{author}{\bibfnamefont{J.}~\bibnamefont{Klug}}, \bibinfo{type}{Diploma
  thesis}, \bibinfo{school}{Uppsala University, Sweden} (\bibinfo{year}{1997}).

\bibitem[{\citenamefont{Conrad et~al.}(2002{\natexlab{a}})\citenamefont{Conrad,
  Botner, Hallgren, and Perez de~los Heros}}]{conrad-1}
\bibinfo{author}{\bibfnamefont{J.}~\bibnamefont{Conrad}},
  \bibinfo{author}{\bibfnamefont{O.}~\bibnamefont{Botner}},
  \bibinfo{author}{\bibfnamefont{A.}~\bibnamefont{Hallgren}}, \bibnamefont{and}
  \bibinfo{author}{\bibfnamefont{C.}~\bibnamefont{Perez de~los Heros}},
  \bibinfo{journal}{Submitted to Phys. Rev. D.}
  (\bibinfo{year}{2002}{\natexlab{a}}), \eprint{arXiv:hep-ex/0202013}.

\bibitem[{\citenamefont{Conrad et~al.}(2002{\natexlab{b}})\citenamefont{Conrad,
  Botner, Hallgren, and Perez de~los Heros}}]{conrad-2}
\bibinfo{author}{\bibfnamefont{J.}~\bibnamefont{Conrad}},
  \bibinfo{author}{\bibfnamefont{O.}~\bibnamefont{Botner}},
  \bibinfo{author}{\bibfnamefont{A.}~\bibnamefont{Hallgren}}, \bibnamefont{and}
  \bibinfo{author}{\bibfnamefont{C.}~\bibnamefont{Perez de~los Heros}},
  \bibinfo{journal}{Proc. of Advanced Statistical Techniques in Particle
  Physics, Durham}  (\bibinfo{year}{2002}{\natexlab{b}}),
  \eprint{arXiv:hep-ex/0206034}.

\bibitem[{\citenamefont{Feldman and Cousins}(1998)}]{feldman}
\bibinfo{author}{\bibfnamefont{G.}~\bibnamefont{Feldman}} \bibnamefont{and}
  \bibinfo{author}{\bibfnamefont{R.}~\bibnamefont{Cousins}},
  \bibinfo{journal}{Phys. Rev. D} \textbf{\bibinfo{volume}{57}},
  \bibinfo{pages}{3873} (\bibinfo{year}{1998}).

\bibitem[{\citenamefont{Learned and Mannheim}(2000)}]{learned-review}
\bibinfo{author}{\bibfnamefont{J.}~\bibnamefont{Learned}} \bibnamefont{and}
  \bibinfo{author}{\bibfnamefont{K.}~\bibnamefont{Mannheim}},
  \bibinfo{journal}{Ann. Rev. Nucl. Part. Sci.} \textbf{\bibinfo{volume}{50}},
  \bibinfo{pages}{679} (\bibinfo{year}{2000}).

\bibitem[{\citenamefont{V.Balkanov et~al.}(2001)}]{venice01:baikal}
\bibinfo{author}{\bibnamefont{V.Balkanov}} \bibnamefont{et~al.}, in
  \emph{\bibinfo{booktitle}{Proc. 9th Int. Workshop on Neutrino Telescopes}}
  (\bibinfo{address}{Venice, Italy}, \bibinfo{year}{2001}),
  vol.~\bibinfo{volume}{II}, p. \bibinfo{pages}{591},
  \eprint{arXiv:astro-ph/0105269}.

\bibitem[{\citenamefont{Balkanov et~al.}(2000)}]{baikal:astropart}
\bibinfo{author}{\bibfnamefont{V.}~\bibnamefont{Balkanov}}
  \bibnamefont{et~al.}, \bibinfo{journal}{Astro. Part. Phys.}
  \textbf{\bibinfo{volume}{14}}, \bibinfo{pages}{61} (\bibinfo{year}{2000}).

\bibitem[{\citenamefont{Rhode et~al.}(1994)}]{frejus}
\bibinfo{author}{\bibfnamefont{W.}~\bibnamefont{Rhode}} \bibnamefont{et~al.},
  \bibinfo{journal}{Astro. Part. Phys.} \textbf{\bibinfo{volume}{4}},
  \bibinfo{pages}{217} (\bibinfo{year}{1994}).

\bibitem[{\citenamefont{Perrone et~al.}(2001)}]{icrc01:macro}
\bibinfo{author}{\bibfnamefont{L.}~\bibnamefont{Perrone}} \bibnamefont{et~al.},
  in \emph{\bibinfo{booktitle}{Proc. 27$^{th}$ Int. Cosmic Ray Conf.}}
  (\bibinfo{address}{Hamburg, Germany}, \bibinfo{year}{2001}), p.
  \bibinfo{pages}{1073}.

\bibitem[{\citenamefont{Dzhilkibaev}(2002)}]{jan:baikal}
\bibinfo{author}{\bibfnamefont{J.}~\bibnamefont{Dzhilkibaev}}
  (\bibinfo{year}{2002}), \bibinfo{note}{private communication. Baikal
  Collaboration}.

\end{thebibliography}

\end{document}